\documentclass[aps,prc,nofootinbib,amsmath,amssymb, onecolumn]{revtex4-2}
\usepackage{graphicx}
\usepackage{dcolumn}
\usepackage{bm}
\usepackage{color}
\usepackage{mathrsfs}
\usepackage{comment}
\usepackage{mathtools}
\usepackage{eurosym}
\usepackage{amsmath,amssymb}
\usepackage{graphicx}
\usepackage{dcolumn}
\usepackage{bm}
\usepackage{color}
\usepackage{multirow}

\usepackage{float}

\usepackage{array}
\usepackage{dcolumn}
\usepackage{booktabs}
\usepackage[detect-all]{siunitx}
\usepackage{makecell}
\usepackage{xcolor}
\usepackage{yfonts}
\newcolumntype{P}[1]{>{\centering\arraybackslash}p{#1}}
\begin{document}
\title{Magnetoexcitons in transition metal dichalcogenides monolayers, bilayers, and van der Waals heterostructures}
\author{Roman Ya. Kezerashvili$^{1,2}$ and Anastasia Spiridonova$^{1,2}\thanks{%
E-mail contact:}$  }
\affiliation{$^{1}$The Graduate School and University Center, The City University of New
York, New York, NY 10016, USA\\
$^{2}$New York City College of Technology, The City University of New York, Brooklyn, NY 11201, USA
}
\date{\today}
\begin{abstract}
We study direct and indirect magnetoexcitons in Rydberg states in monolayers and heterostructures of transition metal dichalcogenices (TMDCs) in an external magnetic field, applied perpendicular to the monolayer or heterostructures in the framework of a Mott-Wannier model of excitons. We calculate binding energies of magnetoexcitons for the Rydberg states 1$s$, 2$s$, 3$s$, and 4$s$ by numerical integration of the Schr\"{o}dinger equation using the Rytova-Keldysh potential for direct magnetoexcitons and both the Rytova-Keldysh and Coulomb potentials for indirect magnetoexcitons. The latter allows understanding the role of screening in TMDCs heterostructures. We report the magnetic field energy contribution to the binding energies and diamagnetic coefficients (DMCs) for direct and indirect magnetoexcitons. The tunability of the energy contribution of direct and indirect magnetoexcitons by the magnetic field is demonstrated. It is shown that binding energies and  DMCs of indirect magnetoexcitons can be manipulated by the number of hBN layers. Therefore, our study raises the possibility of controlling the binding energies of direct and indirect magnetoexcitons in TMDC monolayers, bilayers, and van der Waals heterostructures using magnetic field and opens an additional degree of freedom to tailor the binding energies and DMCs for heterostructures by varying the number of hBN sheets between TMDC layers. The calculations of the binding energies and DMCs of indirect magnetoexcitons in TMDC heterostructures can be compared with the experimental results when they are available.
\end{abstract}
\keywords{}
\maketitle
\vspace{-10mm}
\section{Introduction}
\vspace{-4mm}
Transition metal dichalcogenices (TMDCs) present a great interest for application in nano- and quantum devices at room temperature owing to Dirac cone \citep{Cheng2014,Lee2014,Peng2015,Novoselov2016,Marian2017,Li2017,Iannaccone2018,Xue2018,Tang2019, Marega2020}, direct gap at nonequivalent points $K/K'$  \citep{Mak2010,Splendiani2010,Tongay2012}, and strong-spin orbit interaction that lifts degeneracy at valence and conduction bands \cite{Wang2018,Kosmider2013,Zeng2013}. In design of electronic devices, TMDCs monolayers, TMDC/TMDC, and TMDC/insulator/TMDC heterostructures can be used. Moreover, TMDCs present a great interest for fundamental research. TMDCs heterostructure can support high-temperature quantum Bose gases of indirect excitons \citep{Fogler2014,Berman2016,Calman2018,Calman2020}, superfluidity of electron-hole \citep{Berman2019,Donck2020} and dipolar excitons \citep{Berman2017},
and superconductivity \cite{Cotlet2016}.

\indent TMDCs are semiconductors that have a chemical formula MX$_2$ where M is a transition metal atom (Mo or W) and X is a chalcogen atom (S, Se, or Te). The comprehensive review of excitonic complexes in TMDCs monolayers is given in Refs. \citep{Manzeli2017,Wang2018,Kezerashvili2019}. Excitons in monolayers have a direct energy gap. However, TMDCs heterostructures made of monolayers have indirect gaps. There are two possible arrangements of double-layer heterostructures: bilayer TMDC-TMDC \citep{Wickramaratne2014,Rivera2014,Palummo2015,Chen2016,Nagler2017, Lihet2017,Rivera2016,Rivera2018,Calman2018, Donckhet2018,Kunstmann2018,Ovesen2019,Calman2020} and TMDC-insulator-TMDC \cite{Fogler2014, Calman2016,Donckhet2018}. For both arrangements, there are two types of heterostructures: type-I and type-II. Type-I heterostructure has a direct band gap alignment, and type-II has a staggered band alignment \citep{Rivera2014,Chen2016,Nagler2017,Calman2018,Calman2020}. However, stacking order \cite{Terrones2013, He2014} or application of strain \citep{Amin2015} can be used to induce an indirect gap into a direct gap. In addition, the application of the perpendicular electric field controlled by voltage modifies the band structure, so it is advantageous for electrons and holes to reside in different TMDC layers \cite{Fogler2014, Calman2018,Jauregui2019} and makes indirect excitons more preferable than direct excitons. However, the application of the electric field can lead to dissociation of indirect excitons \cite{Kamban2020}. Nevertheless, the extended lifetime of indirect excitons \citep{Rivera2014,Palummo2015,Miller2017,Rivera2018} compared to direct excitons presents great interest for the design of electronic devices.

\indent The history of studying excitons in semiconductors in the magnetic field extends back for 60 years. Starting from Elliot and Loudon \cite{Elliott1960} and Hasegawa and Howard \cite{Hasegawa1961}, who developed the theory of the Mott exciton in the strong magnetic field followed by authors of Refs. \cite{Shinada1965,Gorkov1967,Akimoto1967}, who studied the physics of Mott magnetoexcitons. For direct magnetoexcitons in TMDCs monolayers, the binding energies of Rydberg states of direct excitons are reported in Refs. \citep{Kylanpaa2015,Stier2016,Zipfel2018,Stier2018,Liu2019, Gor2019,Goldstein2020}, the Zeeman (ZM) shift has been considered in Refs. \cite{Aivazian2014,Srivastava2014,Ludwig2014,Macneill2014,Stier2016,Plechinger2016,Rybkovskiy2017,DonckDM2018, Stier2018,Koperski2018,Chen2019,Gor2019,Liu2019,Xuan2020}, and the diamagnetic (DM) shift was addressed in Refs. \citep{Luckert2010,Aivazian2014,Macneill2014, Choi2015,Stier2016,Plechinger2016,Stier2018, DonckDM2018, donckexc2018, Zipfel2018,Han2018, Chen2019,Liu2019,Gor2019, Spiridonova}. For indirect excitons in double-layer heterostructures Rydberg states binding energies are reported in \citep{He2014,Rivera2014,Palummo2015,Chen2016,Hu2016,Cadiz2017,Gerber2018,Donckhet2018,Rivera2018,Brunetti2018,Kamban2020,Calman2020}, and Zeeman shift is addressed in Refs. \citep{Rivera2016,Lindlau2018,Arora2018,Seyler2019,Wang2020,Wozniak2020}.\\
\indent Historically, the electromagnetic interaction between the electron and hole was described by Coulomb potential ($V_{C}$) for both direct and indirect excitons \citep{Elliott1960,Shinada1965,Gorkov1967,Akimoto1967,Shinada1970,Lozovik1978,Herold1981,MacDonald1986,Stafford1990,Lozovik1997}. However, as has been shown, the Rytova-Keldysh (RK) potential ($V_{RK}$) \citep{Rytova,Keldysh} is the appropriate potential to describe the interaction of electron and hole in two-dimensional (2D) configuration space since the RK potential takes into account the effects of the dielectric environment and the 2D confinement. As a result, currently, both the RK \cite{Kezerashvili2019,Liu2019,Stier2016,Chen2019,Rybkovskiy2017,Gor2019,donckexc2018, Zipfel2018,Goldstein2020,Robert2018,Chernikov2014,Mayers2015,Kidd2016} and Coulomb potentials are used to describe the direct exciton. For indirect excitons, the Coulomb potential is used. However, in a few cases, the RK potential is used as well \citep{Danovich2018,Kamban2020}. We do the same thing: we use both the RK and Coulomb potentials for indirect excitons to investigate the importance of using the accurate potential that describes an interaction between the hole and electron.\\
\indent  Magnetoexcitons in TMDCs monolayers have brought the considerable recent interest with respect to magnetic field tuning than other 2D materials, since they preserve time-reversal symmetry with excitons formed at $K$ and $K'$ points at the boundary of the Brillouin zone, which restricts valley polarization. On one side, the contribution of the magnetic field to the Rydberg states is small, however the tunability of the binding energy of magnetoexcitons brings unique potential for controlling novel device applications in optoelectronics. On the other side, as we demonstrate in this work, the realization of van der Waals
heterostructures consisting of stacked 2D layers, where indirect magnetoexcitons formed, can be significantly controlled by engineering numbers of hBN monolayer, in addition to be tuned by the external magnetic field.

\indent In this paper, we study the effects of the external magnetic field on the binding energies of  1$s$, 2$s$, 3$s$, and 4$s$ Rydberg states of direct and indirect magnetoexcitons in TMDCs. $A$ and $B$ direct magnetoexcitons are considered in freestanding (FS) and encapsulated by the hexagonal boron nitride (hBN) MX$_{2}$ monolayers, and $A$ and $B$ indirect magnetoexcitons are considered in bilayer MX$_{2}$-MX$_{2}$ and MX$_{2}$-hBN-MX$_{2}$. In our study to find eigenfunctions and eigenenergies for indirect magnetoexcitons, a two-particle Schr\"{o}dinger equation is solved numerically using the Rytova-Keldysh and Coulomb potentials to calculate binding energies and diamagnetic coefficients. We demonstrate strong sensitivity of the energy contributions from the external magnetic field on the type of the potential. The first time diamagnetic coefficients (DMCs) for bilayer MX$_{2}$-MX$_{2}$ and  MX$_{2}$-hBN-MX$_{2}$ heterostructures are calculated and reported. Our study raises the possibility to control the binding energies of direct and indirect magnetoexcitons in monolayer MX$_{2}$, bilayer MX$_{2}$-MX$_{2}$, and MX$_{2}$-hBN-MX$_{2}$ heterostructures using a magnetic field and opens an additional degree of freedom to tailor the binding energies and DMCs for heterostructures by varying the number of hBN sheets between TMDC layers.

The paper is organized in the following way. In Sec. \ref{theory} we introduce the theoretical formalism for the description of Mott-Wannier magnetoexcitons in freestanding and encapsulated MX$_2$ monolayers, and in TMDC heterostructures and discuss electrostatic interactions that form direct and indirect magnetoexcitons. In Sec. \ref{results} we focus on the analysis and discussion of the results. In particular, contributions from the external magnetic field to binding energies of magnetoexcitons in MX$_{2}$ monolayers, MX$_{2}$-MX$_{2}$ bilayers, and  MX$_{2}$-hBN-MX$_{2}$ heterostructures are presented and discussed in Secs. \ref{results:monolayer} and \ref{results:bilayer}, respectively. Sec. \ref{diam_coef} is devoted to the calculations of diamagnetic coefficients for direct and indirect magnetoexcitons. Finally, conclusion follows in Sec. \ref{conclusion}.
\vspace{-6mm}
\section{Theoretical Formalism}\label{theory}
\vspace{-4mm}
In this section, we provide the theoretical formalism for describing the Mott-Wannier magnetoexciton in 2D materials and present the energy
contribution from the external magnetic field to the Rydberg states binding energy of magnetoexcitons. Electrostatic interactions that due to a non-local dielectric screening of an electron-hole interaction strongly modify the electrostatic Coulomb potential and form direct magnetoexcitons in TMDC monolayers and indirect magnitoexcitons in MX$_{2}$-MX$_{2}$ bilayers and MX$_{2}$-hBN-MX$_{2}$ van der Waals heterostructure are presented.
\vspace{-7mm}
\subsection{Mott-Wannier magnetoexciton in 2D materials}
\vspace{-4mm}
Our approach is an effective mass model of
excitons and we are using this model for the approximation of the solid state Hamiltonian for an interacting electron and hole. Therefore, our theory is
based on two-body excitonic Hamiltonians in
the effective mass approximation with screened interactions
appropriate for TMDC materials. So, let's introduce the equation for the description of Mott-Wannier excitons in the external magnetic field. We employ the
effective mass model for two charged point-like particles in
two dimensions. In general, electrostatically-bound electrons and holes in the external magnetic field form magnetoexcitons.
To find the eigenfunctions and eigenenergies of a 2D magnetoexciton in
TMDCs monolayer, bilayer MX$_{2}$-MX$_{2}$, and MX$_{2}$-hBN-MX$_{2}$ heterostructure in the external magnetic field, we write the Schr\"{o}dinger
equation for an interacting electron and hole \cite{Herold1981} $(\hbar=c=1)$:
\begin{equation}
\bigg[ \frac{1}{2m_{e}}\Big( -i\nabla _{e}+e\mathbf{A(r}_{e})\Big) ^{2}+%
\frac{1}{2m_{h}}\Big( -i\nabla _{h}-e\mathbf{A(r}_{h})\Big) ^{2}+V\left(
r_{e},r_{h}\right) \bigg] \psi \left( \mathbf{r}_{e},\mathbf{r}_{h}\right)
=E\psi \left( \mathbf{r}_{e},\mathbf{r}_{h}\right) ,  \label{eq:ehschro}
\end{equation}%
where $e$ is the charge of the electron, the indices $e$ and $h$ are referring to the electron and hole,
respectively,  $\mathbf{r}_{e}$ and $\mathbf{r}_{h}$ are 2D coordinates, $m_{e}$ and $m_{h}$ are the masses of charged particles, $\mathbf{A(r}_{e/h})=\mathbf{%
B\times r}_{e/h}/2$ is a gauge vector potential, and $V\left(
r_{e},r_{h}\right) $ is the potential of interaction between the electron and
hole confined in 2D space. In three-dimensional (3D) homogeneous dielectric environments, the electron-hole interaction is
described by the Coulomb potential. However, in 2D monolayer this interaction has to be modified because of the reduced dimensionality and screening effects. The original derivation of two charged particles  interaction in 2D space was given by Rytova \cite{Rytova} and a decade later was independently obtained by Keldysh \cite{Keldysh}. The potential is called the Rytova-Keldysh (RK) potential.
For almost a decade, the RK potential has been used to describe
electrostatic interaction between charge carriers of few-body complexes in TMDCs, phosphorene, and Xenes monolayers (See \cite{Kezerashvili2019} and references herein).

Following Refs. \cite{Elliott1960, Gorkov1967, Shinada1965, Akimoto1967, Herold1981, Lozovik1997}, in
Eq. (\ref{eq:ehschro}) we introduce the coordinate of the center-of-mass $%
\mathbf{R}=\frac{m_{e}\mathbf{r}_{e}+m_{h}\mathbf{r}_{h}}{M}$, where $M=m_{e}+m_{h}$ is  the total mass of the system, and the
relative motion coordinate $\mathbf{r}=\mathbf{r}_{e}-\mathbf{r}_{h}$ and consider the magnetic field
pointing in $z$-direction that is perpendicular to the monolayer where the
exciton is located, $\mathbf{B}=\mathbf{B}(0,0,B)$. After performing the standard
procedure for the coordinate transformation to the center-of-mass, Eq. (\ref{eq:ehschro}) becomes:
\begin{eqnarray}
\bigg [ &-&\frac{ 1}{2M}\frac{{\partial }^{2}}{\partial \bm{R}^{2}}-\frac{%
1}{2\mu }\frac{{\partial }^{2}}{\partial \bm{r}^{2}}+\frac{e^{2}}{%
8\mu }(\bm{B}\times \bm{R})^{2}+\frac{e^{2} \mu ^2}{8}\left( \frac{1}{m_e ^3}+\frac{1}{m_h ^3}\right) (\bm{B}\times \bm{r})^{2}-\frac{%
i e}{2M}(\bm{B}\times \bm{r})\cdot \frac{\partial }{\partial \bm{R}}
\notag \\
&&-\frac{i e}{2\mu }(\bm{B}\times \bm{R})\cdot \frac{\partial }{%
\partial \bm{r}}-\frac{i e\gamma }{2\mu }(\bm{B}\times \bm{r})\cdot
\frac{\partial }{\partial \bm{r}}+\frac{e^{2}\gamma }{4\mu }(\bm{B}%
\times \bm{R})\cdot (\bm{B}\times \bm{r})+V(\bm{r}) \bigg] \psi (\bm{R},\bm{r})=E\psi (\bm{R},\bm{r}),  \label{eq:HRr}
\end{eqnarray}%
where $\gamma =\frac{m_{h}-m_{e}}{m_{h}+m_{e}}$ and $\mu =\frac{m_{e}m_{h}}{%
m_{e}+m_{h}}$ is the reduced mass.
Equation (\ref{eq:HRr}) is written for the case when masses of the electron and hole are different: $m_{h}\neq m_{e}$, which is the case in TMDCs. The term $(\bm{B}\times \bm{r})\cdot\frac{\partial }{\partial \bm{r}}=\bm{B}\cdot \bm{L}=0$ since we consider Rydberg states: 1$s$, 2$s$, 3$s$, and 4$s$, for which $l=0$, $m_l=0$.

Following Refs. \cite%
{Gorkov1967,Lozovik1997,Herold1981}, we introduce an operator $\hat{\bm{P}}$:
\begin{equation}
\hat{\bm{P}}=-i \nabla _{\bm{R}}-\frac{e}{2}(\bm{B}\times \bm{r}).
\label{eq:poperator}
\end{equation}%
It is easy to check that $\hat{\bm{P}}$ commutes with the Hamiltonian in Eq. (\ref{eq:ehschro}), therefore, it has the
same eigenfunction as Eq. (\ref{eq:ehschro}). Thus, one can write the
wave function for the exciton in the magnetic field as \cite{Gorkov1967, Lozovik1997}%
:
\begin{equation}
\psi (\bm{R},\bm{r})=e^{\left[ i\bm{R}\cdot (\bm{P}+\frac{e}{2%
}\bm{B}\times \bm{r})\right] } e^{\frac{1}{2}i \gamma \bm{r}\cdot \bm{P}}\Phi (\bm{r}-\tilde{\bm{\rho}}_0),  \label{eq:wavefun}
\end{equation}%
where the notation
$\tilde{\bm{\rho}}_0=\frac{1}{eB^2}(\bm{B}\times \bm{P})$ is introduced. After substituting the
wave function $\psi (\bm{R},\bm{r})$ in Eq. (\ref{eq:HRr}), the Schr\"{o}dinger equation for the relative
motion of the electron and hole reads:
\begin{equation}
\left[ \frac{\bm{P}^2}{2M} + \frac{e}{4\mu}\bm{P}\cdot (\bm{
B \times \bm{r}})-\frac{1}{2\mu}\frac{{\partial }^{2}}{\partial \bm{r}%
^{2}}+\frac{e^{2}}{8\mu}(\bm{B}\times \bm{r})^{2}-\frac{ie\gamma}{2\mu}(%
\bm{B}\times \bm{r})\cdot \frac{\partial }{\partial \bm{r}}-\frac{i\gamma}{2\mu}\bm{P}\frac{\partial }{\partial \bm{r}}+V(r)\right]
\Phi(\bm{r}) =E\Phi(\bm{r}).   \label{Relmotion}
\end{equation}%
Finally, after separating the angular variable in (\ref{Relmotion}), the equation with zero center-of-mass momentum reads \cite{MacDonald1986, Lozovik1997}:

\begin{equation}
\left[ -\frac{1}{2\mu}\frac{{\partial }^{2}}{\partial r^{2}}-\frac{1}{2\mu}\frac{1}{r}\frac{{\partial }}{\partial r}+\frac{%
e^{2}}{8 \mu}(\bm{B}\times \bm{r})^{2}+V(r)\right] \Phi (r)=E\Phi (r).  \label{eq:finsch}
\end{equation}%
Equation (\ref{eq:finsch}) describes the Mott--Wannier magnetoexciton in Rydberg
optical states in 2D materials. This equation has a long history of the solution in the
case of the electron-hole Coulomb interaction \cite{Elliott1960,
Gorkov1967,Shinada1965,Akimoto1967,Shinada1970,Lozovik1978,Herold1981,MacDonald1986,Stafford1990,
Lozovik1997}. However, we solve Eq. (\ref{eq:finsch}) using $V_{RK}$ for direct magnetoexcitons. For indirect magnetoexcitons, both $V_{RK}$ and $V_{C}$ potentials are used to investigate the importance of using correct potential describing electron-hole interactions. Note that Eq. (\ref{eq:finsch}) does not explicitly contain any spin- or valley-dependent Zeeman terms.

To find binding energies, we numerically solve Eq. (\ref{eq:finsch}) by using the code implemented in Ref. \citep{Brunetti2018} which was successfully modified and used to calculate binding energies of magnetoexcitons in TMDCs monolayers \cite{Spiridonova} and Xenes heterostructures {\cite{Kezerashvili2021}}. The method is based on using the finite element method implemented in Wolfram Mathematica in the NDEigensystem function. The code was modified in a way that the Schr\"{o}dinger equation explicitly contains $\frac{e^2}{8\mu}(\textbf{B}\times \textbf{r})^2$ term. To check the code, we use the input parameters from respective papers listed below and calculate the binding energies of direct and indirect excitons. The code reproduces theoretical binding energies of excitons in TMDCs monolayers reported in Refs. \citep{donckexc2018,Kylanpaa2015,Berkelbach2013}, obtained in the framework of the stochastic variational, the path integral Monte-Carlo and variational methods, respectively, within 5\%. The experimental binding energies reported in Refs. \citep{Stier2018,Gor2019,Liu2019} are reproduced within 7\%. Related to the bilayer system composed from two different TMDC monolayers, we reproduce theoretical binding energies, obtained using a numerical method based on exterior complex scaling given in Ref. \cite{Kamban2020}, within 3\% when the parameter $\kappa$ is varied between 1 and 5. The theoretical binding energies reported in Ref. \cite{Danovich2018}, obtained using quantum Monte Carlo method, are reproduced within 10\%. It is worth noting that in the above papers for the bilayer system, the RK potential is used to described interactions between the electron and hole.

The energy contribution from the magnetic field and DMCs have been considered so far only for direct excitons in TMDCs monolayers where the RK potential is used. In Ref. \cite{Liu2019} the external magnetic field is treated as a small perturbation. In Ref. \cite{Stier2018} Eq. (\ref{eq:finsch}) is solved on a grid, i.e. an unknown function is introduced in order to split the second-order differential equation into two first-order equations. In Ref. \cite{Gor2019} Eq. (\ref{eq:finsch}) is solved numerically. In Ref. \cite{donckexc2018} the stochastic variational method with a correlated Gaussian basis is used to calculated binding energies. They provide values of binding energies of 1$s$ state at different values of the magnetic field. We reproduce their values when we use their parameters in our code.

By calculating the magnetoexciton energy of Rydberg states at different values of the magnetic field, we can find the energy contribution from the magnetic field to the binding energy of a magnetoexciton in the following way:
\begin{equation}
\Delta E= |E(B)-E_0|.\label{eq:contr}
\end{equation}
In Eq. (\ref{eq:contr}) $E_0$ is the exciton binding energy when the magnetic field is absent and is calculated with respect to the two-body threshold \citep{Landau1977,Walck,Berkelbach2013,donckexc2018} while $E(B)$ is the magnetoexciton energy at some value of the magnetic field.

In literature, the Zeeman and diamagnetic shifts of excitons in TMDCs monolayers are treated in the same way as ZM and DM shifts of excitons in quantum wells and quantum dots \citep{Rogers1986,Nash1989,Walck,Erdmann2006,Godoy2006,Kim2009, Abbarchi2010,Bree2012, Brodbeck2017}. The first step is to write a magnetoexciton energy using Taylor series \cite{Walck}:
\begin{equation}
E(B) = E_0 + \gamma _1 B+ \gamma _2 B^2 +...\quad . \label{eq:taylor}
\end{equation}
In Eq. (\ref{eq:taylor}) the second and third terms are the ZM and DM shifts, respectively. When the magnetic field's energy contribution is small compared to the binding energy in the absence of the magnetic field, we can use the first three terms of Taylor series to describe the magnetoexciton binding energy \citep{Walck, Godoy2006, Abbarchi2010,Chen2019}. In other words, the following condition $E_0>|E(B)-E_0|$ is met. However, when $E_0\sim|E(B)-E_0|$, higher-order terms in Eq. (\ref{eq:taylor}) need to be considered.

As stated before, the treatment of the ZM and DM shifts of magnetoexcitons in TMDCs monolayers follows the same procedure applied for quantum dots and quantum wells. The second step in describing ZM and DM shifts is to define them by connecting shifts to terms of Taylor series. The Zeeman shift is defined as $\gamma _1 B=-\mu_B g B$ \citep{Bree2012,Godoy2006,Abbarchi2010,Aivazian2014,Macneill2014,Rybkovskiy2017,Koperski2018,Chen2019}, where $g$ is the effective $g$ factor and $\mu_B$ is Bohr magneton. Experimentally, the ZM shift is defined as energy difference between $K$/$K'$ points i.e. $-g\mu_B B=E(K)-E(K')$ \citep{Kim2009,Abbarchi2010,Stier2016,Plechinger2016,Rybkovskiy2017,Stier2018,Koperski2018,Gor2019,Liu2019,Chen2019}. The diamagnetic shift is defined as $\gamma_2 B^2 =\frac{e^2}{8\mu}\langle r^2 \rangle B^2$ \citep{Rogers1986,Nash1989,Walck,Bree2012,Erdmann2006,Godoy2006,Kim2009, Abbarchi2010, Brodbeck2017, Stier2016, Stier2018, Gor2019, Liu2019,Chen2019}, where $\langle r^2 \rangle$ is the expectation value of $r^2$ over the exciton envelope wave function. In our work, we follow notation used for magnetoexciton diamagnetic coefficient in TMDCs monolayers i.e. $\gamma_2 \equiv \sigma$ \cite{Macneill2014,Stier2016, Stier2018,Koperski2018, Liu2019, Gor2019}. According to Refs. \citep{Rogers1986,Nash1989,Walck,Bree2012,Erdmann2006,Godoy2006,Kim2009, Abbarchi2010, Brodbeck2017, Stier2016, Stier2018,Koperski2018, Gor2019, Liu2019} the DMC for a carrier in a semiconductor is defined as $\sigma =\frac{e^2}{8\mu}\langle r^2 \rangle$. Therefore, by calculating DMCs, exciton radius, reduced mass, and dielectric properties of a material can be obtained \cite{Walck, Stier2016,Stier2018, Liu2019, Gor2019}. Experimentally, the DM shift is defined as $\sigma B^2=\frac{E(K)+E(K')}{2}$ \citep{Godoy2006,Brodbeck2017,Abbarchi2010, Liu2019, Aivazian2014, Chen2019}.
\begin{figure}[b]
\begin{centering}
\includegraphics[width=8.0cm]{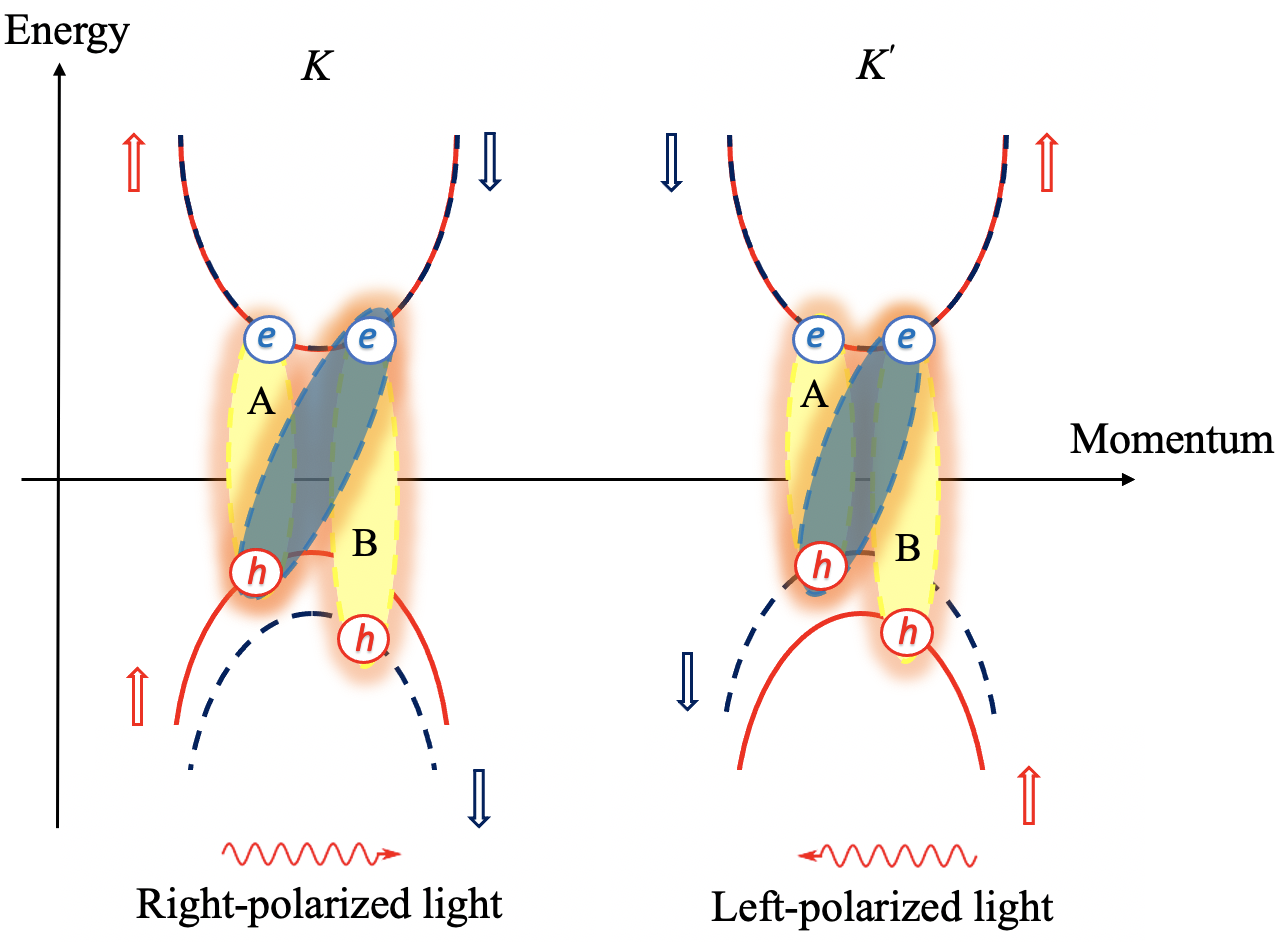}
\caption{The schematic band structure and electronic dispersions in the TMDC monolayer for bright and dark excitons in the $K$
and $K'$ valleys. Spin-up and spin-down bands are denoted by red and blue curves, respectively. The yellow shadowed ovals are the bright
excitons and correspond to the lowest optically induced transition between the bands of the same spin at the $K$ and $K'$ point.
The dark shadowed oval is the spin-forbidden dark exciton (the second one is not shown). The units of the vertical and
horizontal axes are arbitrary. At point $K$ the right circular polarized light couples to both $A$ and $B$ exciton transitions. At
point $K'$ the left circular polarized light couples to $A$ and $B$ excitons.}
\label{exciton_type}
\end{centering}
\end{figure}

In Fig. \ref{exciton_type} we show the schematic formation of bright (optically allowed) direct intravalley $A$ and $B$ excitons in $K$ and $K'$ valleys under
the right- and left-polarized light. We call the excitons that are formed by charge carriers with parallel spins at the conduction and valence bands separated by a small gap as direct $A$ excitons. The excitons formed by charge carriers with parallel spins at the conduction and valence bands separated by a large gap are called direct $B$ excitons \cite{Ramasubramaniam2012,Kylanpaa2015,Stier2016}. There exist two additional exciton types, when charge carriers have antiparallel spins, but these excitons are optically forbidden and called spin-forbidden dark excitons \citep{Echeverry2016,Malic2018}. In the bilayer and MX$_2$-hBN-MX$_2$ structures, the conduction band minimum and valence band maximum reside in two different layers forming the indirect interlayer exciton. There are two possible stacking orders in TMDCs: AA and AB \citep{Pflugradt2014,Yarmohammadi2017,Schneider2019,Zhangadv2020}. In the case of AA stacking interlayer intravalley exciton is formed by the hole from the valence band in $K$ valley of the layer 1 and the electron from the conduction band in $K$ valley of the layer 2 \cite{Zhangadv2020}. Therefore, $\mu$ of direct and indirect excitons are the same. In the case of AB stacking, the lower layer is $180 ^{\circ}$ in plane rotation of the upper layer \cite{Jones2014}. So, band
spins of $K$ valley of the layer 1 are flipped compared to corresponding band spins of $K$ valley of the layer 2. Therefore, the bright interlayer intravalley exciton is formed by the hole with $m_B$ in $K$ valley of the layer 1 and the electron with $m_A$ in $K$ valley of the layer 2 \citep{Arora2017, Horng2018,Gerber2019,Lorchat2021}.
\vspace{-7mm}
\subsection{Magnetoexcitons in TMDCs monolayer}
\begin{figure}[H]

\begin{centering}
\includegraphics[width=14.0cm]{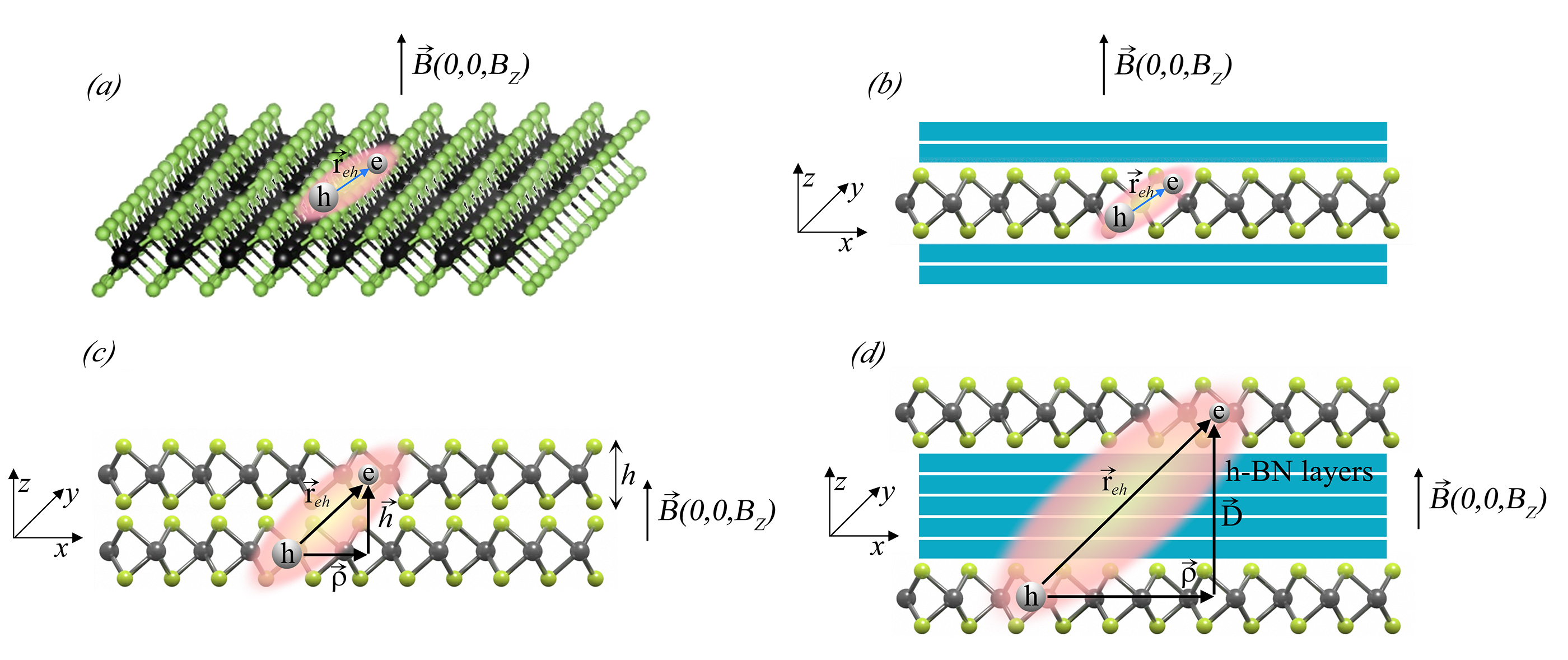}
\caption{(Color online) The schematic illustration of magnetoexcitons in TMDC monolayers and heterostructures. $(a)$ A direct magnetoexciton in a freestanding TMDC monolayer. $(b)$ A direct magnetoexciton in an encapsulated TMDC monolayer. $(c)$ An indirect magnetoexciton in a freestanding bilayer MX$_{2}$-MX$_{2}$ heterostructure. $(d)$ An indirect magnetoexciton in MX$_{2}$-hBN-MX$_{2}$ van der Waals heterostructure.}
\label{fig1M}
\end{centering}
\end{figure}

\vspace{-4mm}
A non-local dielectric screening of an electron-hole interaction strongly modifies the electrostatic
Coulomb potential. It leads to a non-hydrogenic Rydberg series of exciting magnetoexciton states. The Rytova-Keldysh potential has become a prevalent description of the electrostatic interaction of charge carriers in 2D  systems. The RK potential is central
potential, and the interaction between the electron and hole for direct
excitons in an encapsulated TMDC monolayer has the form \cite{Rytova, Keldysh}:
\begin{equation}
V_{RK}(r)=-\frac{\pi ke^{2}}{2\kappa \rho _{0}}\left[ H_{0}\left( \frac{r}{\rho
_{0}}\right) -Y_{0}\left( \frac{r}{\rho _{0}}\right) \right] ,  \label{eq:rk}
\end{equation}%
where $r\equiv r_{eh}=r_{e}-r_{h}$ is the relative coordinate between the electron and
hole. In Eq.~(\ref{eq:rk}), $\kappa
=(\epsilon _{1}+\epsilon _{2})/2$ describes the surrounding dielectric
environment, $\epsilon _{1}$ and $\epsilon _{2}$ are the dielectric
constants below and above the monolayer, $H_{0}$ and $Y_{0}$ are the Struve and Bessel
functions of the second kind, respectively, and $\rho _{0}=2\pi \chi _{2D}/\kappa$ \cite{Berkelbach2013} is the
screening length, where $\chi _{2D}$ is the 2D polarizability of the monolayer, which is given by $\chi_{2D}=h\epsilon /4\pi $ \cite{Keldysh}, where $\epsilon $ is the bulk dielectric constant of the electron/hole containing monolayer and $h$ is the thickness of a TMDC monolayer. At $r < \rho _{0}$ the RK potential has logarithmical dependence and diverges logarithmically at the origin. At
long-range distances, $r > \rho _{0}$, it retains the Coulomb potential behavior. Using  the RK potential for description of excitons in 2D monolayer, one finds a strong dependence of the binding energy on whether the monolayer is suspended in air (Fig. \ref{fig1M}$a$), encapsulated by  hBN (Fig. \ref{fig1M}$b$), or in heterostructures (Figs. \ref{fig1M}$c$ and \ref{fig1M}$d$). It is worth mentioning that the RK potential is obtained for a monolayer. However, a TMDC monolayer consists of three atomic layers: chalcogen-metal-chalcogen sheets. To address this problem, a new potential is derived in Ref. \cite{Dery2018} which considers the three atomic sheets that compose a monolayer of TMDC and can explain the non-hydrogenic Rydberg series of excitons in TMDC monolayer. The Rytova-Keldysh potential can be recovered when considering the strict 2D limit, and in our study, we are using the RK potential.

\vspace{-8mm}
\subsection{Indirect magnetoexcitons in TMDCs heterostructures}
\vspace{-4mm}
Van der Waals attractive forces can hold together layered TMDC materials. Let us now consider the formation of indirect magnetoexcitons in the heterostructure molded by the same type of TMDC monolayers: a bilayer MX$_{2}$-MX$_{2}$ heterostructure, shown in Fig. \ref{fig1M}$c$ and van der Waals MX$_{2}$-hBN-MX$_{2}$ heterostructure, where two TMDC monolayers are
separated by $N$ layers of hBN monolayers as the ones used in most of the
experiments, presented in Fig. \ref{fig1M}$d$. In such structures, the electron and
hole move in the planes of different layers, form an indirect exciton, and have restricted motion between the layers due to the dielectric barrier. The latter is a two-body problem in the restricted 3D space, and the electrostatic electron-hole interaction $V(r)$
could form a bound state, i.e., the indirect
magnetoexciton. Therefore, to determine
the binding energy of the magnetoexciton one must solve a two-body problem in the
restricted 3D space due to the restriction of the motion in $z$-direction. The relative separation $r$ between the
electron and hole can be written in cylindrical coordinates as $\bm{r}=\rho \hat{\bm{\rho}}+D\hat{\bm{z}}$, where $\hat{\bm{\rho}}$ and $\hat{\bm{z}}$ are unit vectors. Writing $r$ in cylindrical coordinates allows us to treat the case of direct excitons in a TMDC
monolayer and spatially indirect excitons in bilayer MX$_{2}$-MX$_{2}$ and MX$_{2}$-hBN-MX$_{2}$ heterostructures on
equal footing. If we set $D=0$ in the latter expression, it becomes a purely
2D equation, with $\rho $ representing the separation between the electron
and hole sharing the same plane. For spatially indirect excitons, the relative distance between the electron and hole is $r=%
\sqrt{\rho ^{2}+D^{2}}$, where $D$ is the distance between the middle of two TMDC layers assuming that the electron and hole reside in the middle of their respective sheets \cite{Kamban2020}.

Thus, for the description of indirect Mott--Wanner magnetoexcitons
one can use Eq. (\ref{eq:finsch}) with interactions
\begin{equation}
V_{RK}(\sqrt{\rho ^{2}+D^{2}})=-\frac{\pi ke^{2}}{2\kappa \rho _{0}}\left[
H_{0}\left( \frac{\sqrt{\rho ^{2}+D^{2}}}{\rho _{0}}\right) -Y_{0}\left(
\frac{\sqrt{\rho ^{2}+D^{2}}}{\rho _{0}}\right) \right]   \label{eq:indkeld}
\end{equation}%
for the RK potential, and
\begin{equation}
V_C \left( \sqrt{\rho ^{2}+D^{2}}\right) =-\frac{ke^{2}}{\kappa \left( \sqrt{\rho
^{2}+D^{2}}\right) }  \label{eq:indcoul}
\end{equation}%
for the Coulomb potential. In Eq.~\eqref{eq:indkeld} $\rho_{0} = \rho^{(1)}_{0} + \rho^{(2)}_{0}$, where $\rho^{(1)}_{0}$ and $\rho^{(2)}_{0}$ are the screening lengths of the first and second monolayer, respectively. Polarizability of both layers needs to be taken into account since the total screening length has increased. If polarizabilities of both layers are not taken into account then binding energies are higher \cite{Kamban2020}. Our calculations show that if polariziblity for only one layer is considered then the binding energies of 1$s$ state increase by about 25-30\%. Equations~\eqref{eq:indkeld} and~\eqref{eq:indcoul}
describe the interaction between the electron located in one and hole in the other parallel TMDC monolayers.
Therefore, one
can obtain the eigenfunctions and eigenenergies of magnetoexcitons by
solving Eq.~ (\ref{eq:finsch}) using the potential~(\ref{eq:rk}) for direct
magnetoexcitons, or using either potential~(\ref%
{eq:indkeld}) or~(\ref{eq:indcoul}) for indirect magnetoexcitons. For indirect magnetoexcitons, we perform calculations using both the RK and Coulomb potentials. This allows a better understanding of the importance of the screening effect in MX$_{2}$-MX$_{2}$ and MX$_{2}$-hBN-MX$_{2}$ heterostructures.

It is worth mentioning that the RK potential was originally formulated as an
explicitly 2D description of the Coulomb interaction.
Nevertheless, there have been recent attempts to apply the RK potential to
indirect excitons in van der Waals heterostructures of 2D
materials such as the TMDCs, phosphorene, and Xenes~\cite%
{Fogler2014,Berman2016,Berman2017b,Berman2017,Brunetti2018,Brunetti2018b,Kezerashvili2021}. The logic behind considering the RK potential for indirect excitons
follows from two considerations: i. the dielectric environment is still
inhomogeneous, just as in the case of the direct exciton --{} when the
interlayer separation $D$ is smaller than, or comparable to, the RK potential
screening length $\rho _{0}$ and the excitonic gyration radius $\sqrt{%
\langle r^{2}\rangle }$, the electron-hole interaction potential must
account for both the TMDC monolayers and the interlayer dielectric, and ii.
as the interlayer separation $D$ becomes larger than $\rho _{0}$, the total
separation between the electron and hole, $r=\sqrt{\rho ^{2}+D^{2}}$,
 becomes much larger than $\rho _{0}$, and, therefore, the RK
potential converges towards the Coulomb potential. Let us emphasize
that we are not claiming definitively that the RK potential provides the
most accurate description of the spatially indirect exciton, hence, we present the extensive comparison with the corresponding results obtained using the
Coulomb potential.

Above, we consider ideal cases when encapsulated TMDC monolayer, bilayer MX$_{2}$-MX$_{2}$, and MX$_{2}$-hBN-MX$_{2}$ heterostructures are fabricated so that
each layered material is stacked without any vacuum or air interlayer gap. In reality, between the layers always exists non-vanishing interlayer gap.
In each layer the field lines of the Coulomb interaction
are screened by the adjacent material, which reduces the single-particle band gap as well as exciton binding energies \cite{Florian2018}. In Ref. \cite{Florian2018}  the authors demonstrate and give a quantitative understanding that the binding energy of excitons and electronic and optical properties are sensitive to the interlayer distances on the atomic scale.

Throughout this paper, we consider the separation between the electron and hole residing in two TMDC monolayers of the  MX$_{2}$-hBN-MX$_{2}$ van der Waals heterostructures in steps of calibrated thickness, $%
l_{\text{hBN}}=0.333~\text{nm}$, corresponding to the thickness of one hBN monolayer: $D=h+Nl_{\text{hBN}}$, where $h$ is the TMDC
monolayer thickness and $N$ is the number of hBN monolayers. For a bilayer  MX$_{2}$-MX$_{2}$ heterostructure $N = 0$ and following Ref. \cite{Kamban2020} the electron and hole reside in the middle of their
respective sheets and $r=\sqrt{\rho ^{2}+h^{2}}$.
\vspace{-6mm}
\section{Results of calculations and discussion}\label{results}
\vspace{-4mm}
We report the energy contribution from the external magnetic field to the binding energies and diamagnetic coefficients of $A$ and $B$ magnetoexcitons in Rydberg optical states 1$s$, 2$s$, 3$s$, and 4$s$, in freestanding (Fig. \ref{fig1M}$a$) and encapsulated (Fig. \ref{fig1M}$b$) WSe$_2$, WS$_2$, MoSe$_2$, and MoS$_2$ monolayers, bilayer MX$_{2}$-MX$_{2}$ (Fig. \ref{fig1M}$c$), and  MX$_{2}$-hBN-MX$_{2}$ heterostructures (Fig. \ref{fig1M}$d$).

The diamagnetic coefficients for the bilayer MX$_{2}$-MX$_{2}$ and  MX$_{2}$-hBN-MX$_{2}$ heterostructures are reported for the first time. We are considering the further knob to tailor the binding energies and diamagnetic coefficients for the MX$_{2}$-hBN-MX$_{2}$ heterostructures by varying the number of hBN sheets between TMDC layers and present corresponding calculations.

In our calculations, we vary the magnetic field in the increment of 1 T in the range from 0 T to 30 T and use input parameters given in Table \ref{table:parameters} presented in Appendix \ref{app:param}. In Table \ref{table:parameters}, the values of $A_{\text{high}}$ and $A_{\text{low}}$ correspond to the parameters found in the literature, which maximize and minimize the $A$ exciton binding energy, respectively. It is highly likely that each parameter's true value for a given material falls somewhere within the given range. Therefore, the true magnitude of the calculated quantities for $A$ excitons studied in this paper lies somewhere
between the calculated values. Values of $B$ correspond to parameters given in the literature for $B$ excitons.

\vspace{-7mm}
\subsection{Contribution from the external magnetic field to binding energies of magnetoexcitons in a monolayer}\label{results:monolayer}
\vspace{-4mm}
\begin{figure}[b]
\begin{centering}
\includegraphics[width=16.0cm]{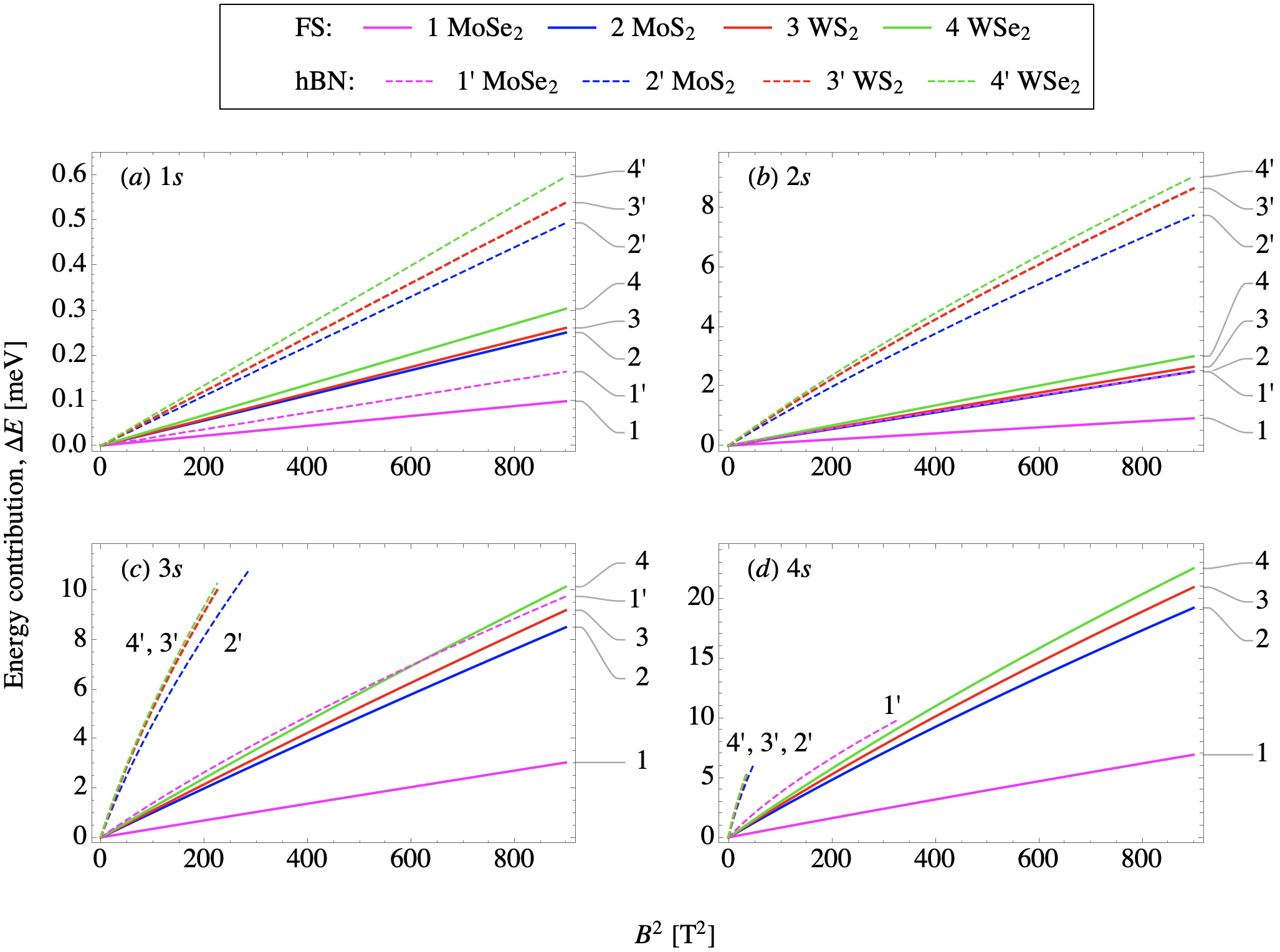}
\caption{Energy contribution from the magnetic field to the binding energies of Rydberg states 1$s$ ($a$), 2$s$ ($b$), 3$s$ ($c$), and 4$s$ ($d$) for direct $A_{\text{low}}$ magnetoexcitons in FS and encapsulated by hBN MX$_2$ monolayers. The solid and dashed curves present results of calculations for freestanding and encapsulated TMDC monolayers, respectively. The magnetic field, where the dissociation of magnetoexcitons in states 3$s$ and 4$s$ occurs, corresponds to
the end of broken curve.}
\label{mon_4_mat}
\end{centering}
\end{figure}
The results of the energy contribution from the magnetic field to the binding energy of Rydberg states of direct $A_{\text{low}}$ magnetoexcitons in freestanding and encapsulated TMDC monolayers are reported in Fig. \ref{mon_4_mat}.
The comparative analysis of the results presented in Fig. \ref{mon_4_mat} shows the following: i. the energy contribution from the magnetic field to the binding energy of the direct magnetoexcitons in FS monolayers of TMDC materials is always less than in encapsulated monolayers, and the difference increases with the increase of the magnetic field; ii. the direct magnetoexcitons in FS monolayers are bound in 1$s$, 2$s$, 3$s$, and 4$s$ states within all considered range of the magnetic field, while magnetoexcitons in encapsulated monolayers dissociated in the states 3$s$ and 4$s$ when the magnetic field is greater than some particular values. These conclusions are related to the direct $A_{\text{low}}$ magnetoexcitons. Interestingly enough that these conclusions stay the same for direct $A_{\text{high}}$ and $B$ magnetoexcitons. However, the $\Delta E$ is systematically smaller for the $A_{\text{high}}$, while $\Delta E$ for $B$ magnetoexcitons lie between  $\Delta E$ for the $A_{\text{low}}$ and $A_{\text{high}}$ magnetoexcitons. Analysis of the results shows that the energy contribution from the external magnetic field to the
magnetoexciton binding energy depends on material parameters. In fact, the binding energy of magnetoexcitons is a function of the reduced mass $\mu$ and polarizability $\chi_{2D}$: the binding energy of a magnetoexciton is bigger for the larger $\mu$ and smaller $\chi_{2D}$, while, when $\mu$ is smaller and $\chi_{2D}$ is larger, $\Delta E$ is bigger.
This pattern qualitatively coincides with previously reported calculations
for magnetoexcitons in WSe$_{2}$ and MoSe$_2$ monolayers \cite{Spiridonova}.

Results of our calculations show three distinct features: i. the energy contribution from the magnetic field to magnetoexcitons in the FS and encapsulated monolayers for the state 1$s$ is more than one order of magnitude smaller than for the states 2$s$, 3$s$ and 4$s$; ii. the magnetoexcitons in FS monolayers are bound in states 1$s$, 2$s$, 3$s$ and 4$s$ when the magnetic field varies up to 30 T; iii. magnetoexcitons in encapsulated monolayers are dissociated in states 3$s$ and 4$s$ at some values of the magnetic field while staying bound in states 1$s$ and 2$s$ at higher magnetic field values.

\begin{figure}[h!]
\begin{centering}
 \includegraphics[width=160mm]{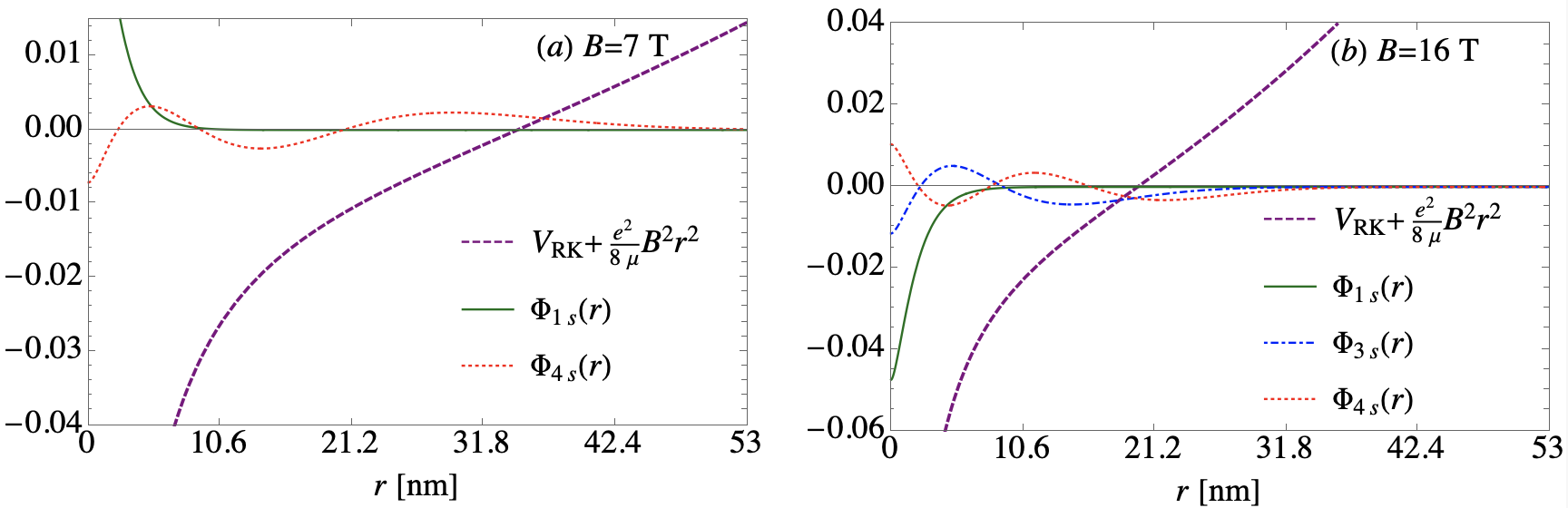}
\caption{Dependencies of the 1$s$, 3$s$, 4$s$ states wavefunctions and total potential energy on the electron-hole distance $r$. The graphs show the behavior of the total potential and wave functions of states 1$s$, 3$s$, and 4$s$ of $A_{\text{low}}$ magnetoexciton in WSe$_2$ monolayer encapsulated by hBN at different values of the magnetic field.
The magnetoexciton dissociates in states 3$s$ and 4$s$ at 16 T and 7 T, respectively. The graphs demonstrate the behavior of $\Phi_{3s}(r)$ and $\Phi_{4s}(r)$ that dissociate at 16 T and 7 T, respectively, with respect to the bound $\Phi_{1s}(r)$ and the corresponding $V_{RK}(r)+\frac{e^2}{8\mu}B^2 r^2$ potential.}  \label{potential}
\vspace{-6mm}
\end{centering}
\end{figure}

Let us address the a dissociation of the magnetoexcitons in encapsulated monolayers in states 3$s$ and 4$s$ at some values of the magnetic field.
To understand the dissociation behavior of magnetoexcitons in 3$s$ and 4$s$ states in TMDC monolayer, following Refs. \citep{Schiff1968,Davydov,Sakurai2017} we examine the characteristic behavior of the wavefunction and the total interaction potential of the electron-hole system in the external magnetic field. In Fig. \ref{potential} is shown the behavior of wavefunctions for different states and the corresponding total potential $V_{RK}+\frac{e^2}{8\mu}B^2 r^2$. The total potential of the system, $V_{RK}(r)+\frac{e^2}{8\mu}B^2 r^2$, and the wavefunctions, $\Phi_{1s}(r)$, $\Phi_{3s}(r)$ and $\Phi_{4s}(r)$ are plotted as a function of $r$. The total potential is given at two different values of the magnetic field when the dissociation of the 3$s$ and 4$s$ states occurs. As can be seen from Fig. \ref{potential}, the total potential becomes positive as the magnetic field increases. The the wavefunction for the bound 1$s$ state is localized, while the wavefunctions for the 3$s$ and 4$s$ states are delocalized at $B = 16$ T and $B = 7$ T, respectively.

\vspace{-7mm}
\subsection{Contribution from the external magnetic field to binding energies of magnetoexcitons in bilayer MX$_{2}$-MX$_{2}$ and MX$_{2}$-hBN-MX$_{2}$ heterostructures} \label{results:bilayer}
\begin{figure}[t]
\begin{centering}
\includegraphics[width=16.0cm]{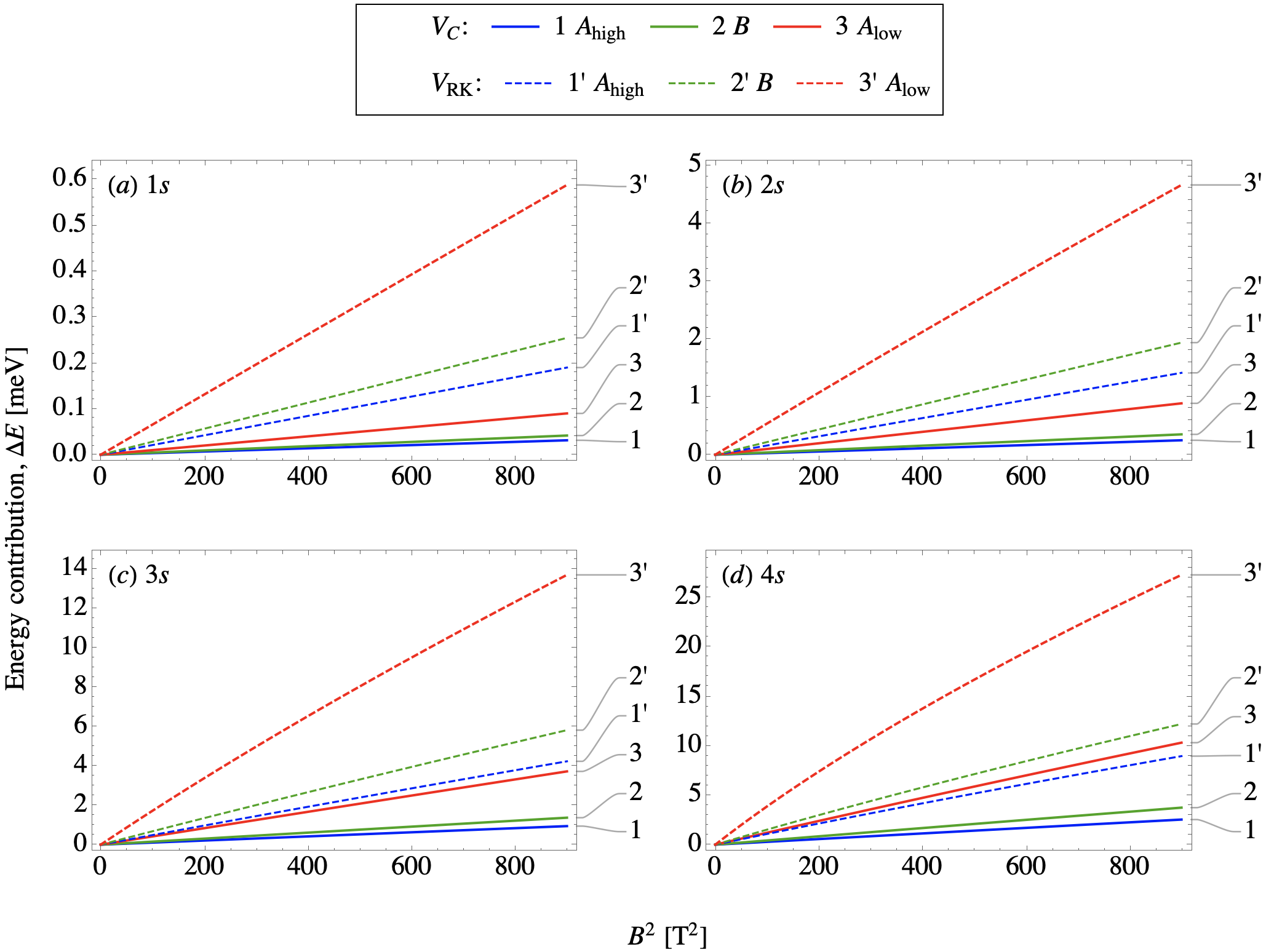}
\caption{Energy contribution from the magnetic field to the binding energies of Rydberg states 1$s$ ($a$), 2$s$ ($b$), 3$s$ ($c$), and 4$s$ ($d$) for indirect magnetoexcitons in MoS$_2$-MoS$_2$ bilayer. The binding energy is calculated with $V_{RK}$ (dashed lines) and $V_C $ (solid lines) potentials. Data are plotted for $A_{\text{high}}$, $A_{\text{low}}$, and $B$ magnetoexcitons.}
\label{mos_bilayer_vac}
\end{centering}
\end{figure}

\vspace{-4mm}
First, we consider bilayer system. We compare the contributions to binding energies of indirect magnetoexcitons due to the external magnetic field calculated using the RK and Coulomb potentials. In Fig. \ref{mos_bilayer_vac}, as a representative case, are shown the energy contributions from the magnetic field to the binding energies of indirect magnetoexcitons in bilayer MoS$_2$-MoS$_2$ for Rydberg states 1$s$ (\ref{mos_bilayer_vac}$a$), 2$s$ (\ref{mos_bilayer_vac}$b$), 3$s$ (\ref{mos_bilayer_vac}$c$), and 4$s$ (\ref{mos_bilayer_vac}$d$). The binding energies are calculated using  $V_{RK}$ and $V_C$ potentials. 
Data are plotted for $A_{\text{high}}$, $A_{\text{low}}$, and $B$ magnetoexcitons. The comparison shows that  i. $\Delta E_{RK} > \Delta E_{C}$ and
the difference increases with the magnetic field increase; ii. results obtained with both potentials satisfy the following inequality: $\Delta E_{A_{\text{high}}} < \Delta E_{B} < \Delta E_{A_\text{low}}$. Most importantly, it is worth to mention that the contribution from the magnetic field is systematically bigger for the bilayer MoS$_{2}$-MoS$_{2}$ than for the FS MoS$_{2}$ monolayer for all states when calculations are performed with $V_{RK}$. In other words, $\Delta E_{A_{\text{low}}}$
for indirect magnetoexcitons in MoS$_{2}$-MoS$_{2}$ bilayer is greater than $\Delta E_{A_{\text{low}}}$ for direct magnetoexcitons in FS MoS$_{2}$ monolayer. Similar patterns are observed for other bilayers constituent from MoSe$_{2}$, WS$_{2}$ and WSe$_{2}$ monolayers, respectively. It is interesting to note that for the bilayer MX$_{2}$-MX$_{2}$ system the energy
contribution from the magnetic field for the state 1$s$ is more than one order of magnitude smaller than for the states 2$s$, 3$s$ and 4$s$ and the states do not dissociate in the range of the varying magnetic field.

\begin{figure}[t]
\begin{centering}
\includegraphics[width=16.0cm]{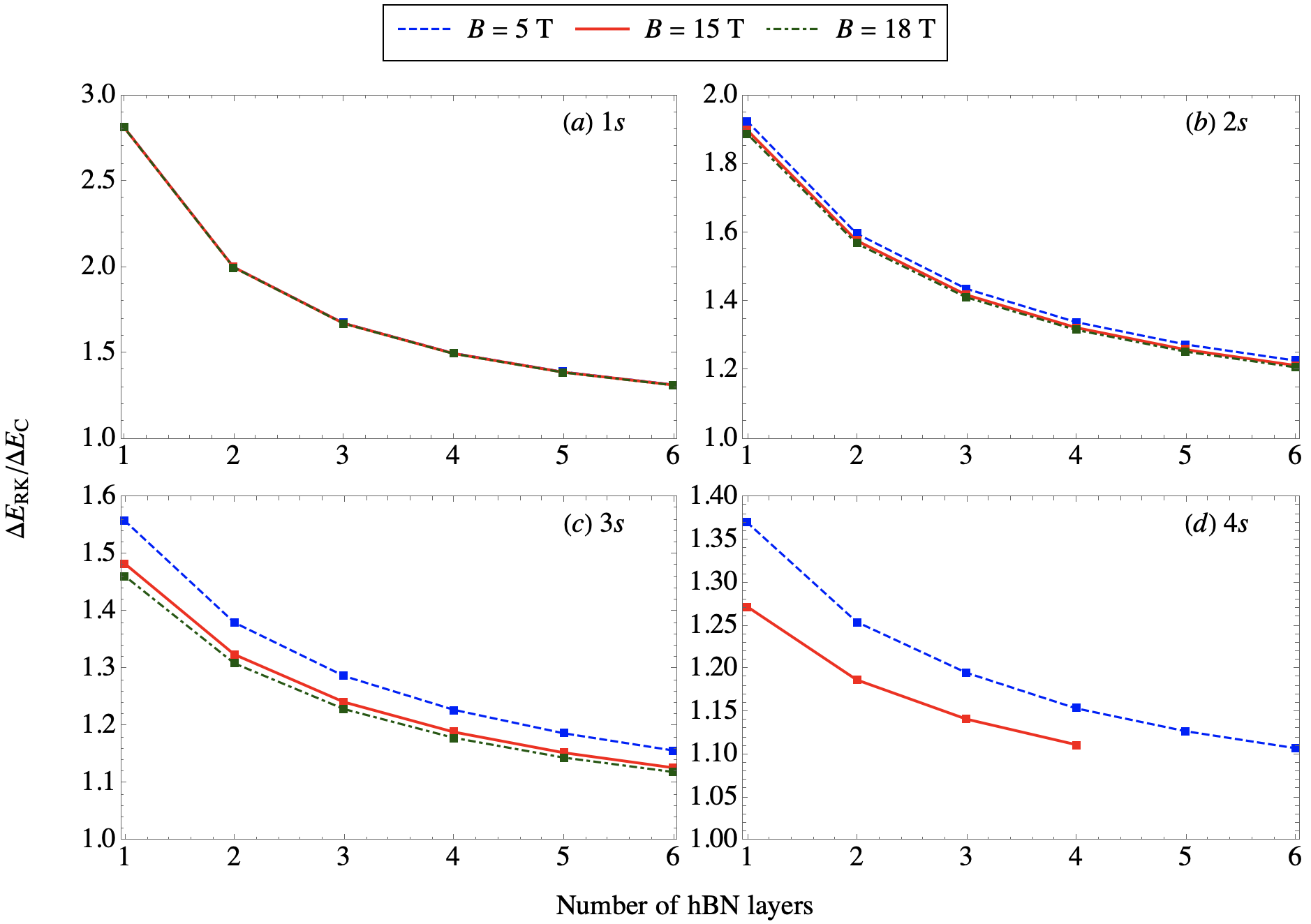}
\caption{Ratio of $\Delta E_{RK}/\Delta E_{C}$ for $A_{\text{high}}$ magnetoexcitons in MoS$_2$-hBN-MoS$_2$ heterostructure for Rydberg states 1$s$ $(a)$, 2$s$ $(b)$, 3$s$ $(c)$, and 4$s$ $(d)$.}
\label{mos_ratio}
\end{centering}
\end{figure}

Next, we consider van der Waals heterostructure. Since the binding energies for excitons in TMDCs monolayers are reported in numerous publications, for example in Refs. \citep{Berkelbach2013,Kylanpaa2015,Stier2016, Stier2018,donckexc2018,Gor2019, Liu2019}. As the first step, we calculate the binding energies of indirect excitons in MX$_2$-hBN-MX$_2$ heterostructures using RK and Coulomb potentials in the absence of the magnetic field. The results of the calculations for 1$s$ Rydberg state are presented in Table \ref{table:binding_energies}, Appendix \ref{app:rydberg}. At the next step, we consider indirect magnetoexcitons in MoS$_{2}$-hBN-MoS$_{2}$ heterostructure. In Fig. \ref{mos_ratio} are presented the results of calculations for the dependence of the ratio $\Delta E_{RK}/\Delta E_{C}$ on the magnetic field and number of the hBN layers separating two parallel TMDC layers. Calculations are performed for $A_{\text{high}}$ magnetoexcitons. The ratio weakly depends on the magnetic field and converges towards 1 when the number of hBN layers increases. The RK potential always gives the larger contribution than the Coulomb potential. Such a result is understandable because the potential $V_{RK}(\sqrt{\rho ^{2}+D^{2}})$ converges to  $V_C \left( \sqrt{\rho ^{2}+D^{2}}\right)$ when $D$ increases. From the known asymptotic properties of the Struve and Bessel functions \cite{Abramowitz, Ryzhik}, it is easy to show that $\displaystyle{\lim_{\rho \rightarrow 0}\frac{V_{RK}}{V_{C}}=\frac{\pi D}{2\rho _{0}}\left[H_{0}(\frac{D}{{\rho _{0}}})-Y_{0}(\frac{D}{{\rho _{0}}})\right]}$. An additional comment to Fig. \ref{mos_ratio}, in the 4$s$ state at $B$ = 15 T, when $N$ = 5, and at $B$ = 18 T the magnetoexcitons are unbound by the RK potential, and due to the dissociation of magnetoexcitons this ratio is zero.

\begin{figure}[t]
\begin{tabular}{cc}

\textit{(a)} 1$s$ &  \textit{(b)} 2$s$ \\
\
  \includegraphics[width=80mm]{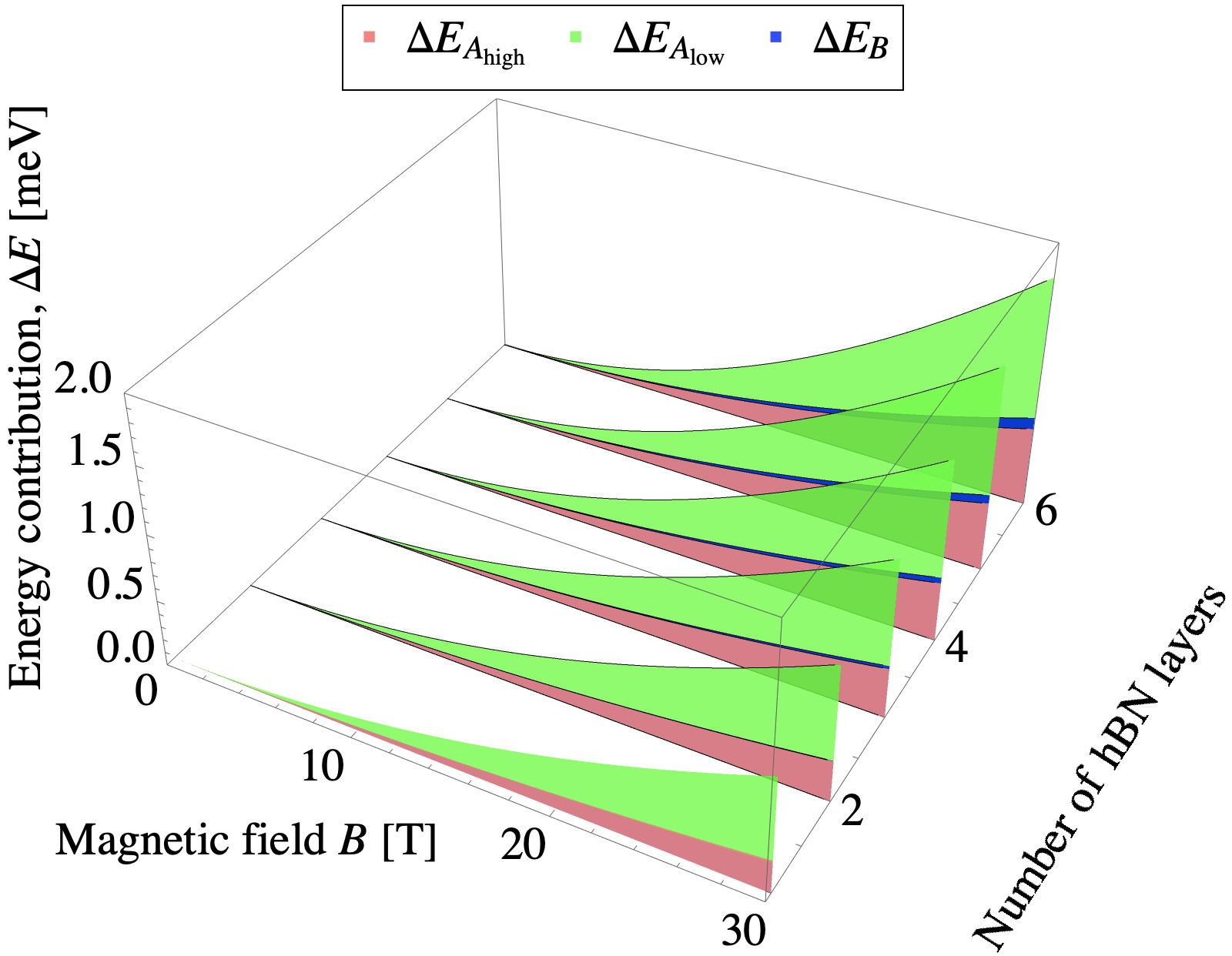} &   \includegraphics[width=80mm]{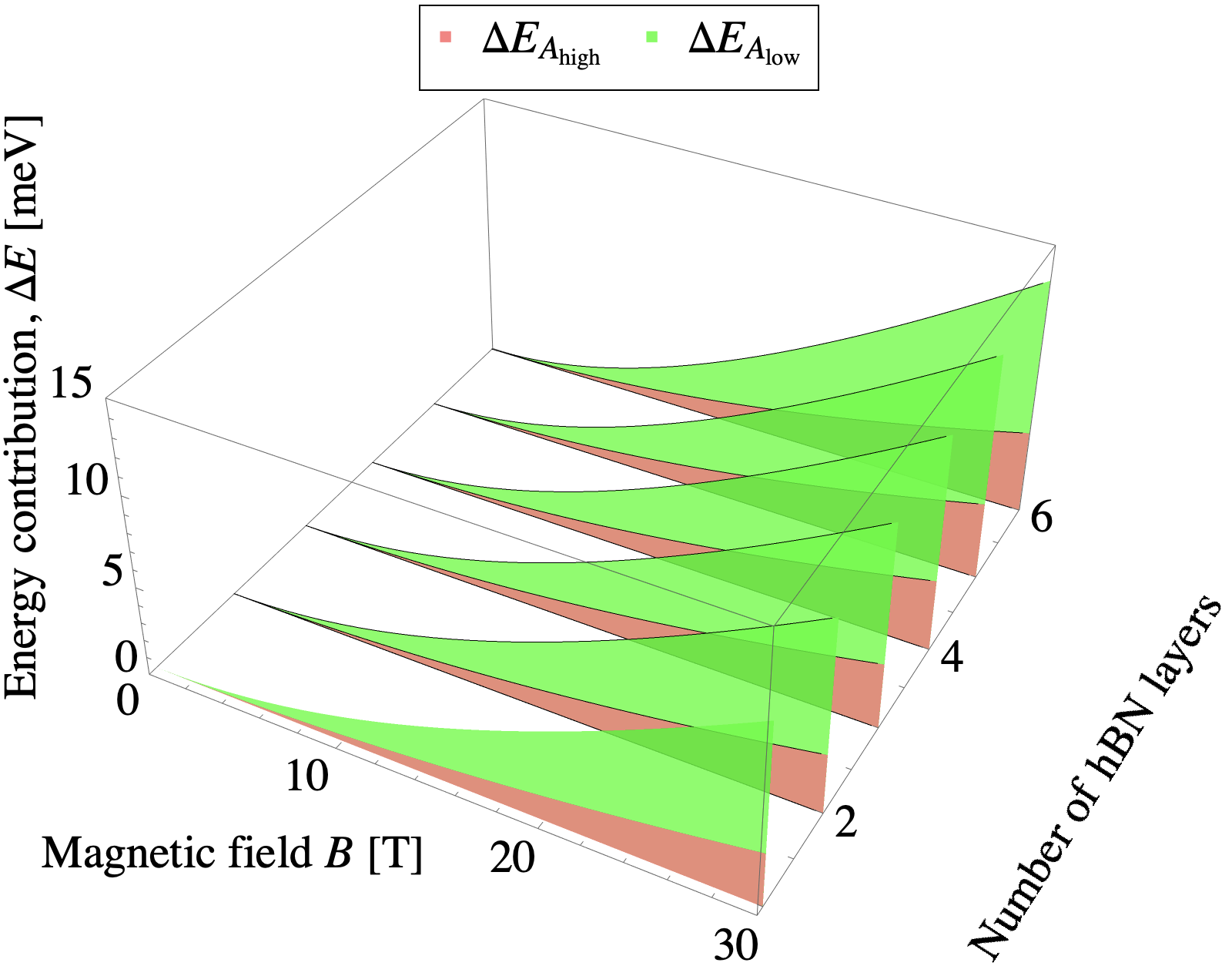} \\[6pt]

\textit{(c)} 3$s$ &  \textit{(d)} 4$s$ \\
\
  \includegraphics[width=80mm]{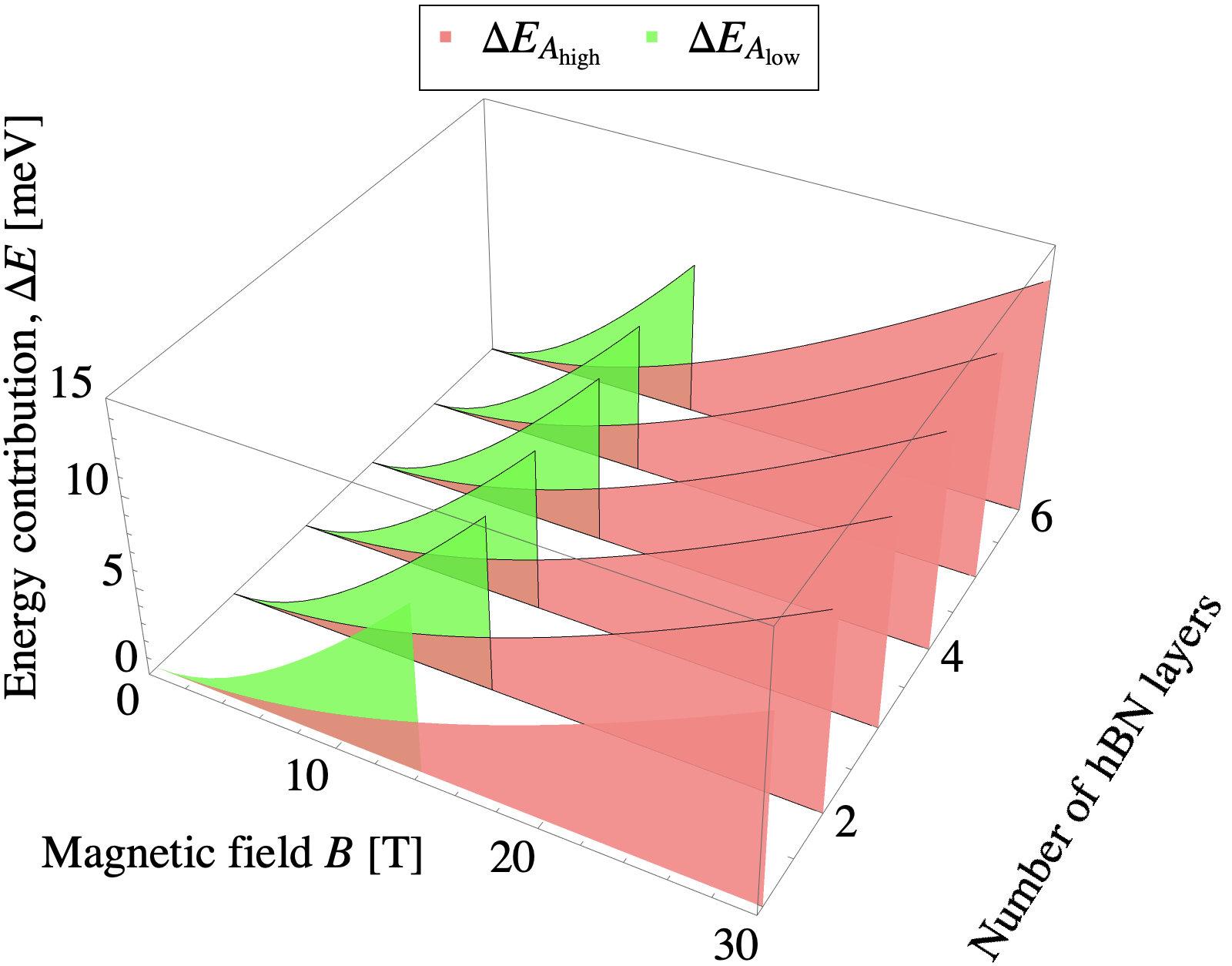} &   \includegraphics[width=80mm]{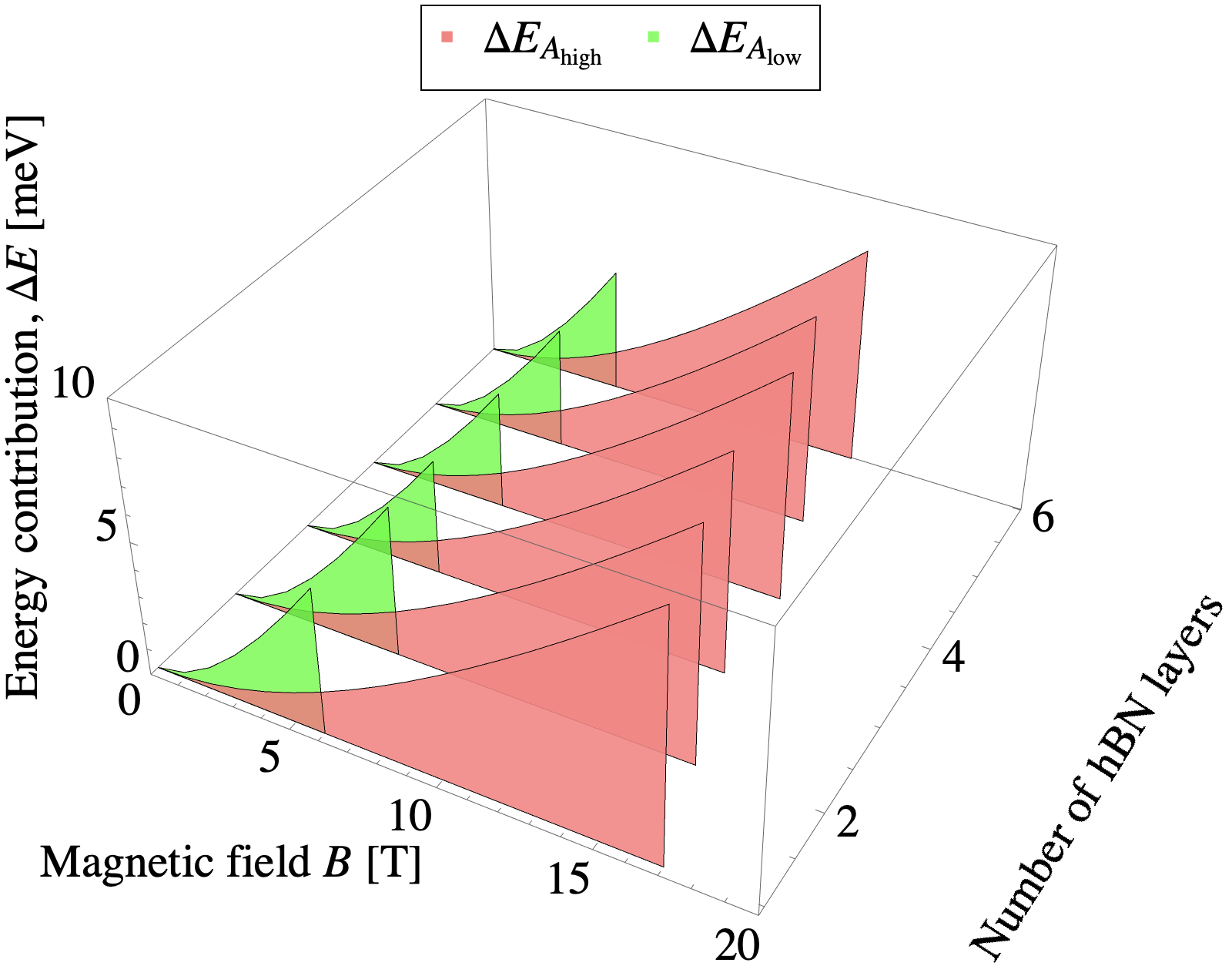} \\[6pt]

\end{tabular}
\caption{The energy contribution from the magnetic field to the binding energies of indirect magnetoexcitons in MoS$_{2}$-hBN-MoS$_{2}$ heterostructure for Rydberg states 1$s$ $(a)$, 2$s$ $(b)$, 3$s$ $(c)$, and 4$s$ $(d)$. Calculations are performed using the Rytova-Keldysh potential. The surface edge tips for the 3$s$ and 4$s$ states
correspond to the magnetic field where the dissociation of the magnetoexciton occurs.}  \label{mos_rk}
\vspace{-6mm}
\end{figure}

The dependencies of the energy contribution to the binding energy of magnetoexcitons Rydberg states for indirect magnetoexcitons on the magnetic field and number of hBN layers for  MoS$_{2}$-hBN-MoS$_{2}$ heterostructure are reported in Fig. \ref{mos_rk}. For 1$s$ and 2$s$ states the energy contribution increases with the increase of the external magnetic field and number of the hBN layers. Indirect magnetoexcitons dissociate in state 3$s$ when $B > 11-15$ T, and in 4$s$ state, when $B > 5$ T. In Fig. \ref{mos_rk}$a$, it is shown the energy contribution for the $B$ exciton always falls between $\Delta E_{A_{\text{high}}}$ and $\Delta E_{A_{\text{low}}}$: $\Delta E_{A_{\text{high}}} < \Delta E_{B} < \Delta E_{A_{\text{low}}}$. The same holds for the states 2$s$, 3$s$ and 4$s$, but not shown in Figs. \ref{mos_rk}$b$, \ref{mos_rk}$c$, and \ref{mos_rk}$d$. It is worth mentioning that Fig. \ref{mos_rk} shows a representative case for indirect magnetoexcitons in  MX$_{2}$-hBN-MX$_{2}$ heterostructure. The same kind patterns and quantitative dependencies we obtained for the WSe$_{2}$-hBN-WSe$_{2}$ heterostructures shown in Appendix \ref{app:het}, Fig. \ref{wse_rk}. Data for WS$_2$-hBN-WS$_2$ and MoSe$_2$-hBN-MoSe$_2$ are not shown. The differences occur only in the energy contribution magnitudes, which varies within up to 40\% for MoSe$_{2}$-hBN-MoSe$_{2}$ for the maximum value, and indirect magnetoexcitons do not dissociate in the 3$s$ state and dissociate at larger values of the magnetic field in 4$s$ state. The tangerine-based heterostructures have the same kind of patterns as are observed for the molybdenum-based heterostructures. In the same Appendix \ref{app:het}, we present in Fig. \ref{mose_low}, as a representative case, the comparison of the energy contribution to the binding energy of indirect magnetoexcitons for MoSe$_{2}$-hBN-MoSe$_{2}$ heterostructure obtained by solving the Schr\"{o}dinger equation with the $V_{RK}$ and $V_{C}$ potentials. The analysis shows that the $V_{RK}$ potential gives higher contributions to the binding energy than the Coulomb potential for all states, and these contributions increase with the increase of the magnetic field, and as the number of hBN layers increase contributions converge as is also shown in Fig. \ref{mos_ratio}. The same qualitative patterns are observed in other heterostructures.

\vspace{-8mm}
\subsection{Diamagnetic coefficients}\label{diam_coef}
\vspace{-4mm}
The diamagnetic coefficients $\sigma_{A_{\text{high/low}}}$ for $A$ and $\sigma_B$ for $B$ direct magnetoexcitons in encapsulated MX$_2$ monolayers obtained in the framework of our approach are reported in Ref. \cite{Spiridonova}.
Here, we report DMCs for direct magnetoexcitons in FS MX$_2$ monolayers and indirect magnetoexcitons in MX$_2$-MX$_2$ bilayers and MX$_2$-hBN-MX$_2$ heterostructures in Tables \ref{table:coefficients_mon_vac/hbn} - \ref{table:coefficients_Se} and Table \ref{table:coefficients_S}, Appendix \ref{app:coef}. \\
\indent The DMCs of direct magnetoexcitons in FS monolayers along with the DMCs for magnetoexcitons in encapsulated WSe$_2$, WS$_2$, MoSe$_2$, and MoS$_2$ are presented in Table \ref{table:coefficients_mon_vac/hbn}. Calculations are performed using the $V_{RK}$ potential. The DMCs $\sigma_{A_{\text{high}}} < \sigma_B < \sigma_{A_{\text{low}}}$ for all states. For the 1$s$ state $\sigma_{\text{hBN}} > \sigma_{\text{FS}}$ and the ratio $\sigma_{\text{hBN}}/\sigma_{\text{FS}}$ varies from 1.6 to 2.6, depending on monolayer parameters. For the 2$s$ state this ratio significantly exceeds 2.6. Therefore, the screening effect in encapsulated monolayers substantially increases the value of DMCs. In addition, as can be seen from Table \ref{table:coefficients_mon_vac/hbn}  in states 3$s$ and 4$s$ $E_0 \sim |E(B)-E_0|$ and Eq. (\ref{eq:taylor}) cannot be applied to extract $\sigma$.

\begin{table}
\caption{The diamagnetic coefficients  $\sigma_{\text{FS}}$ of direct magnetoexcitons in FS MX$_{2}$ monolayers along with DMCs $\sigma_{\text{hBN}}$ in encapsulated MX$_{2}$ monolayers reported in Ref. \cite{Spiridonova}. DMCs are given in $\mu$eV/T$^2$ and obtained when $R^2 = 0.9998$ for the linear regression model.}

\label{table:coefficients_mon_vac/hbn}
\begin{center}
  \sisetup{table-format=2.4}

 \begin{tabular}{P{0.8cm}P{1.5cm}P{1.5cm}P{1.5cm}P{2cm}|P{1.5cm}P{1.5cm}P{2cm}|P{1.5cm}}
\hline\hline
\multicolumn{1}{c}{} &
 \multirow{2}{*}{Exciton} &
\multicolumn{3}{c|}{1$s$}&
\multicolumn{3}{c|}{2$s$}&
\multicolumn{1}{c}{3$s$}
 \\ \cline{3-9}

 \multicolumn{1}{c}{} &
 \multicolumn{1}{c}{} &
\multicolumn{1}{c}{$\sigma_{\text{FS}}$}  &
\multicolumn{1}{c}{$\sigma_{\text{hBN}}$ \cite{Spiridonova}}  &
  \multicolumn{1}{c|}{$\sigma_{\text{hBN}}/\sigma_{\text{FS}}$}  &
  \multicolumn{1}{c}{$\sigma_{\text{FS}}$}  &
  \multicolumn{1}{c}{$\sigma_{\text{hBN}}$ \cite{Spiridonova}}  &
  \multicolumn{1}{c|}{$\sigma_{\text{hBN}}/\sigma_{\text{FS}}$}  &
  \multicolumn{1}{c}{$\sigma_{\text{FS}}$}
   \\ \cline{2-9}
\multirow{3}{*}{WSe$_2$} & $A_{\text{high}}$ & 0.09 & 0.16 & 1.78 & 0.89 & 2.58 & 2.90 & 3.00 \\
						 & $A_{\text{low}}$ & 0.34 & 0.66 & 1.94 & 3.34 & 10.01 & 3.00 \\
						 & $B$              & 0.28 & 0.57 & 2.04 & 2.79 & 8.84 &  3.17 \\ \hline
\multirow{3}{*}{WS$_2$} & $A_{\text{high}}$ & 0.11 & 0.21& 1.91 & 1.10  & 3.56 &  3.24 & 3.81 \\
						 & $A_{\text{low}}$ & 0.29 & 0.60 &  2.07 & 2.94 & 9.65 & 3.28 \\
						 & $B$              & 0.28 & 0.61 & 2.18 & 2.82 & 9.84 & 3.45 \\ \hline
\multirow{3}{*}{MoSe$_2$} & $A_{\text{high}}$ & 0.08 & 0.13 & 1.63 & 0.74 & 1.96 & 2.65 & 2.44 \\
						 & $A_{\text{low}}$ & 0.11 & 0.18 & 1.64 & 1.02 & 2.76 & 2.71 & 3.38 \\
						 & $B$              & 0.09 & 0.17 & 1.89 & 0.85 & 2.55 & 3.00 & 2.83  \\ \hline
\multirow{3}{*}{MoS$_2$} & $A_{\text{high}}$ & 0.08 & 0.21 &  2.63 & 0.77 & 3.56 & 4.62 & 2.61 \\
						 & $A_{\text{low}}$ & 0.28 & 0.55 & 1.96 & 2.77 & 8.64 & 3.12 \\
						 & $B$              & 0.11 & 0.22 & 2.00 & 1.08 & 3.50 & 3.24 & 3.68 \\ \hline \hline
\end{tabular}
\end{center}
\end{table}
\begin{table}
\caption{The diamagnetic coefficients of indirect magnetoexcitons in MX$_{2}$-MX$_{2}$ bilayers. $\sigma_{RK}$ and $\sigma_{C}$ are calculated using V$_{RK}$ and V$_C$ potentials, respectively. DMCs are obtained when $R^2 = 0.9998$ for the linear regression model. $\sigma$ is given in $\mu$eV/T$^2$.}
\label{table:coefficients_bilayer}
\begin{center}
  \sisetup{table-format=2.4}

 \begin{tabular}{P{0.8cm}P{1.5cm}P{1cm}P{1cm}P{1cm}|P{1cm}P{1cm}P{1cm}|P{1cm}P{1cm}P{1cm}|P{1cm}P{1cm}P{1cm}}
\hline\hline
\multicolumn{1}{c}{} &
 \multirow{2}{*}{Exciton} &
\multicolumn{3}{c|}{1$s$}&
\multicolumn{3}{c|}{2$s$}&
\multicolumn{3}{c|}{3$s$}&
\multicolumn{3}{c}{4$s$}
 \\ \cline{3-14}

 \multicolumn{1}{c}{} &
 \multicolumn{1}{c}{} &
\multicolumn{1}{c}{$\sigma_{RK}$}  &
 \multicolumn{1}{c}{$\sigma_{C}$} &
  \multicolumn{1}{c|}{$\sigma_{RK}/\sigma_{C}$}  &
  \multicolumn{1}{c}{$\sigma_{RK}$}  &
 \multicolumn{1}{c}{$\sigma_{C}$} &
  \multicolumn{1}{c|}{$\sigma_{RK}/\sigma_{C}$}  &
  \multicolumn{1}{c}{$\sigma_{RK}$}  &
 \multicolumn{1}{c}{$\sigma_{C}$} &
  \multicolumn{1}{c|}{$\sigma_{RK}/\sigma_{C}$}  &
  \multicolumn{1}{c}{$\sigma_{RK}$}  &
 \multicolumn{1}{c}{$\sigma_{C}$} &
  \multicolumn{1}{c}{$\sigma_{RK}/\sigma_{C}$}
   \\ \cline{1-14}
\multirow{3}{*}{WSe$_2$} & $A_{\text{high}}$ & 0.25 & 0.04 & 6.14 & 1.82 & 0.31 & 5.78 & 5.42 & 1.18 & 4.59 &  & 3.18 &   \\
                         & $A_{\text{low}}$  & 0.79 & 0.12 & 6.53 & 6.23 & 1.19 & 5.22 &      & 5.00 &      &  &  \\
                         & $B$               & 0.67 & 0.11 & 6.26 & 5.27 & 1.03 & 5.13 &      & 4.26 &      &  &    \\ \hline

\multirow{3}{*}{WS$_2$} & $A_{\text{high}}$  & 0.29 & 0.05 & 5.67 & 2.20 & 0.43 & 5.10 & 6.69 & 1.69 &      & & 3.95 \\
                         & $A_{\text{low}}$  & 0.68 & 0.11 & 6.00 & 5.50 & 1.15 & 4.49 &      & 4.87 &      &  &   \\
                         & $B$               & 0.65 & 0.11 & 5.74 & 5.28 & 1.15 & 4.60 &      & 4.87 &      &  &   \\ \hline

\multirow{3}{*}{MoSe$_2$} & $A_{\text{high}}$& 0.21 & 0.03 & 6.79 & 1.53 & 0.23 & 6.60 & 4.49 & 0.85 & 5.31 &  & 2.24\\
                         & $A_{\text{low}}$  & 0.28 & 0.04 & 7.00 & 2.06 & 0.31 & 6.59 &      & 1.18 &      &  & 3.17 \\
                         & $B$               & 0.24 & 0.03 & 6.82 & 1.75 & 0.27 & 6.53 &      & 0.99 &      & & 2.65  \\ \hline

\multirow{3}{*}{MoS$_2$} & $A_{\text{high}}$ & 0.21 & 0.03 & 6.05 & 1.57 & 0.28 & 5.58 & 4.72 & 1.05 & 4.52 &  & 2.82 \\
                         & $A_{\text{low}}$  & 0.65 & 0.10 & 6.54 & 5.20 & 0.99 & 5.28 &      & 4.14 &      &   &   \\
                         & $B$               & 0.28 & 0.05 & 6.09 & 2.15 & 0.39 & 5.52 & 6.50 & 1.52 & 4.27 &  & 4.16
   \\ \hline \hline

\end{tabular}
\end{center}
\end{table}

\begin{table}
\caption{The diamagnetic coefficients of indirect magnetoexcitons in the molybdenum-based van der Waals heterostructures. DMCs are calculated using V$_{RK}$ and V$_C$ potentials. The DMCs are obtained when $R^2 = 0.9998$ for the linear regression model. $\sigma$ is given in $\mu$eV/T$^{2}$.}

\label{table:coefficients_Se}
\begin{center}
  \sisetup{table-format=2.4}

 \begin{tabular}{P{1cm}P{1cm}P{1cm}P{1cm}P{1cm}P{1cm}P{1cm}P{1cm}|P{1cm}P{1cm}P{1cm}P{1cm}P{1cm}P{1cm}}
\hline\hline
\multicolumn{2}{c}{}&
\multicolumn{6}{c|}{MoSe$_{2}$-hBN-MoSe$_{2}$}&
\multicolumn{6}{c}{MoS$_{2}$-hBN-MoS$_{2}$}
 \\ \cline{3-14}
 \multirow{2}{*}{State} &
 \multirow{2}{*}{$N$} &
 \multicolumn{2}{c}{$\sigma_{A_{\text{high}}}$} &
  \multicolumn{2}{c}{$\sigma_{A_{\text{low}}}$}  &
  \multicolumn{2}{c|}{$\sigma_B$} &
 \multicolumn{2}{c}{$\sigma_{A_{\text{high}}}$} &
  \multicolumn{2}{c}{$\sigma_{A_{\text{low}}}$}  &
  \multicolumn{2}{c}{$\sigma_B$}
   \\ \cline{3-14}
  &&  $V_{RK}$ & $V_C$ & $V_{RK}$ & $V_C$ & $V_{RK}$ & $V_C$ &$V_{RK}$ & $V_C$ & $V_{RK}$ & $V_C$ & $V_{RK}$ & $V_C$
  \\ \cline{1-14}
    \multirow{6}{*}{1$s$}& 1 & 0.26 & 0.08 & 0.35 & 0.11 & 0.30 & 0.09  & 0.28 & 0.10 & 0.97 & 0.37 & 0.39 & 0.14  \\
    & 2 & 0.32 & 0.14 & 0.43 & 0.19 & 0.36 & 0.16 & 0.35 & 0.17 & 1.15 & 0.58 & 0.48 & 0.24 \\
    & 3 & 0.38 & 0.21 & 0.51 & 0.27 & 0.44 & 0.24 & 0.42 & 0.25 & 1.34 & 0.81 & 0.57 & 0.35 \\
    & 4 & 0.45 & 0.28 & 0.60 & 0.36 & 0.51 & 0.32 & 0.50 & 0.34 & 1.54 & 1.04 & 0.68 & 0.46 \\
    & 5 & 0.52 & 0.35 & 0.69 & 0.46 & 0.59 & 0.40 & 0.59 & 0.43 & 1.75 & 1.27 & 0.79 & 0.57 \\
    & 6 & 0.60 & 0.43 & 0.78 & 0.55 & 0.68 & 0.48 & 0.68 & 0.52 & 1.96 & 1.51 & 0.90 & 0.69 \\ \cline{1-14}
   \multirow{3}{*}{2$s$}& 1 &    & 1.41 &  & 2.00 &  & 1.67 &  & 1.83 &  &  &  & 2.69   \\
   & 2 &  & 1.90 &  & 2.65 &  & 2.23 \\
   & 3 &  & 2.35 &  &  &  &  &
   \\ \hline \hline

\end{tabular}
\end{center}
\end{table}

Results of calculations of DMCs for the molybdenum- and the tungsten-based van der Waals heterostructures are presented in Table \ref{table:coefficients_Se} and Table \ref{table:coefficients_S}, Appendix \ref{app:coef}, respectively. We present results for $\sigma_{A_{\text{high}}}$ and $\sigma_{A_{\text{low}}}$ for indirect $A$ magnetoexcitons obtained with parameters for each material that give the highest and lowest binding energies, $\sigma_{A_{\text{high/low}}}$ for $A$ excitons and $\sigma_B$ for $B$ magnetoexcitons, respectively. For each type of indirect magnetoexciton, we report two sets of $\sigma$: from solution of the Schr\"{o}dinger equation with $V_{RK}$  and $V_{C}$ potentials, respectively. The DMCs for MoS$_{2}$-hBN-MoS$_{2}$ are always a bit higher than for MoSe$_{2}$-hBN-MoSe$_{2}$ heterostructure. The latter is related to the small difference of the gaps between valence and conduction bands for MoS$_{2}$ and MoSe$_{2}$ monolayers. As follows from Table \ref{table:coefficients_S}, Appendix \ref{app:coef}, the same conclusion can be extended to the DMCs for WS$_{2}$-hBN-WS$_{2}$ and WSe$_{2}$-hBN-WSe$_{2}$ heterostructures. However, the DMCs for the tungsten-based heterostructures are significantly larger than for the molybdenum-based heterostructures because the larger difference of the gaps between valence and conduction bands in the tungsten- and molybdenum-based monolayers \cite{Wang2018}.

The data analysis of Tables \ref{table:coefficients_Se} and \ref{table:coefficients_S} shows the following important features for both S$_2$- and Se$_2$ - based heterostructures: i. the DMCs of the $A$ and $B$ magnetoexcitons increase when the number of hBN layers increase; ii. the DMCs obtained using Rytova-Keldysh potential are always bigger than one obtained for the Coulomb potential; iii. for the states 2$s$, 3$s$ and 4$s$ $E_0 \sim |E(B)-E_0|$ and Eq. (\ref{eq:taylor}) cannot be applied to extract $\sigma_{A_{\text{high/low}}}$ and $\sigma_B$ from data; iv. for both the RK and Coulomb potentials $\sigma_{A_{\text{high}}} < \sigma_B < \sigma_{A_{\text{low}}}$. The latter indicates that DMCs are sensitive to the exciton reduced mass and polarizability of material: the smaller reduced mass of exciton and larger polarizability lead to the bigger DMC value.

A distinct feature for the diamagnetic coefficients in bilayer and van der Waals molybdenum-based heterostructures is that $\sigma$ is always smaller than for the tungsten-based heterostructures due to the difference of the valence and conduction bands in MoX$_2$ and WX$_2$ monolayers. The other distinct feature is that the diamagnetic coefficients can be extracted for the 1$s$, 2$s$, 3$s$, and 4$s$ states for the bilayers, while the DMCs exist in the van der Waals heterostructure with up to six hBN layers for 1$s$ and three hBN layers, for 1$s$ and 2$s$ states, respectively. The linear dependence of $\Delta E$ on $B^2$ is invalidated in 3$s$ and 4$s$ states for both the Rytova-Keldysh and Coulomb potentials.



\vspace{-8mm}
\section{Conclusion} \label{conclusion}
\vspace{-4mm}
In the present paper, we have studied energy contributions from the external magnetic field to the binding energies of $A$ and $B$ magnetoexcitons in 1$s$, 2$s$, 3$s$, and 4$s$ Rydberg states in WSe$_2$, WS$_2$, MoSe$_2$, and MoS$_2$ freestanding and encapsulated monolayers, bilayer MX$_{2}$-MX$_{2}$, and  MX$_{2}$-hBN-MX$_{2}$ heterostructures. The first time diamagnetic coefficients for the bilayer MX$_{2}$-MX$_{2}$ and  MX$_{2}$-hBN-MX$_{2}$ heterostructures are calculated and reported. We consider the additional degree of freedom to tailor the binding energies and diamagnetic coefficients for the MX$_{2}$-hBN-MX$_{2}$ heterostructures by varying the number of hBN sheets between TMDC layers. The study is performed for direct and indirect $A$ and $B$ excitons. For the $A$ magnetoexcitons we used two sets of parameters given in literature that provide the highest ($A_{\text{high}}$ exciton) and lowest ($A_{\text{low}}$ exciton) binding energy for excitons. For indirect magnetoexcitons in bilayer and van der Waals heterostructures the dependence of energy contribution from the external magnetic field to the binding energies and DMCs are studied using Rytova-Keldysh and Coulomb potentials.

Our study shows that the energy contributions $\Delta E$ for direct and indirect magnetoexcitons and DMCs are sensitive to the reduced mass of exciton and the screening distance $\rho_{0}$: the smaller reduced mass and larger $\rho_{0}$ lead to the bigger $\Delta E$ and DMCs.

The energy contributions $\Delta E$ for direct magnetoexcitons increase with the increase of the external magnetic field, and $\Delta E$ is more significant for 2$s$, 3$s$ and 4$s$ states.  The DMCs for direct magnetoexcitons in encapsulated monolayers are bigger than in FS monolayers, $\sigma_{\text{hBN}} > \sigma_{\text{FS}}$, and the ratio $\sigma_{\text{hBN}}/\sigma_{\text{FS}}$ varies from 1.6 to 3.6 in states 1\textit{s} and 2\textit{s} depending on monolayer parameters and the state of magnetoexcitons. Therefore, the screening effect in encapsulated monolayers significantly increases the value of DMCs.

For our investigation, a two-particle Schr\"{o}dinger equation for indirect excitons is solved using the Rytova-Keldysh and Coulomb potentials to calculate binding energies and diamagnetic coefficients. We demonstrate strong sensitivity of the energy contributions from the external magnetic field on the type of the potential: $\Delta E_{RK} > \Delta E_{C}$ and
the difference increases with the increase of the magnetic field. For all structures the energy contribution have the following order: $\Delta E_{A_{\text{high}}} < \Delta E_{B} < \Delta E_{A_\text{low}}$. The DMCs for MX$_{2}$-hBN-MX$_{2}$ heterostructures obtained using Rytova-Keldysh potential are always bigger than one obtained for
the Coulomb potential and the DMCs of the $A$ and $B$ magnetoexcitons increase with the increase of the number of
hBN layers. In the considered range of the magnetic field, the direct and indirect magnetoexcitons have stable 1$s$ and 2$s$ Rydberg states. The 3$s$ and 4$s$ states magnetoexcitons dissociate at some values of the magnetic field depending on parameters of material. The 3$s$ and 4$s$ states indirect magnetoexcitons bound by the Coulomb potential dissociate at a higher magnetic field than magnetoexcitons bound by the RK potential.\\
\indent Related to bilayer MX$_{2}$-MX$_{2}$: the magnetic field's contribution to the magnetoexcitons binding energies is systematically bigger for the bilayer than for the FS MX$_{2}$ monolayer for all states when the Rytova-Keldysh potential is used in calculations. The same conclusion is extended to DMCs.\\
\indent A distinct feature for the diamagnetic coefficients in bilayer and van der Waals molybdenum-based heterostructures is that $\sigma$ is always smaller for the tungsten-based heterostructures due to the difference of the valence and conduction bands in MoX$_2$ and WX$_2$ monolayers. The other distinct feature is that within the linear regression model with $R^2 = 0.9998$ the diamagnetic coefficients can be extracted for the 1$s$, 2$s$, 3$s$, and 4$s$ states for the bilayers, while the DMCs exist in the van der Waals heterostructure with up to six hBN layers for 1$s$ and three hBN layers, for 1$s$ and 2$s$ states, respectively. The linear dependence of $\Delta E$ on $B^2$ is invalidated in 3$s$, and 4$s$ states for both the Rytova-Keldysh and Coulomb potentials.\\
\indent Finally, our results raise the possibility of controlling the binding energies of direct and indirect magnitoexcitons in monolayer MX$_{2}$, bilayer MX$_{2}$-MX$_{2}$ and MX$_{2}$-hBN-MX$_{2}$ heterostructures  using the external magnetic field and open the additional degree of freedom to tailor the binding energies and diamagnetic coefficients for the MX$_{2}$-hBN-MX$_{2}$ heterostructures by varying the number of hBN sheets between TMDC layers.
\\

\textbf{Acknowledgments.}
We are grateful to the reviewers for valuable suggestions and comments. This work is supported by the U.S. Department of Defense under Grant No. W911NF1810433 and PSC-CUNY Award No. 62261-00 50.

\appendix

\setcounter{table}{0} \renewcommand{\thetable}{A.\arabic{table}}

\section{Input parameters} \label{app:param}
\vspace{-10mm}

Below are given the input parameters for $A$ and $B$ excitons.
\vspace{-20mm}
\begin{table}[H]
\caption{Parameters for the $A$ and $B$ excitons. For the $A$ exciton for each material two
values of $\protect\mu$ and $\protect\chi_{2D}$ are given. The values of $A_{\text{high}}$ and $A_{\text{low}}$ correspond to the parameters found in the literature, which maximize and minimize the $A$ exciton binding energy, respectively The average dielectric constant, $%
\protect\kappa=\frac{\protect\varepsilon_1+\protect\varepsilon_2}{2} =4.89$,
is taken from Ref. \protect\cite{Fogler2014}. $\protect\kappa$ is the same for
all materials. $\protect\mu$ is in units of electron mass, $m_{0}$.The 2D polarizability $%
\protect\chi_{2D}$ and TMDC monolayer thickness $h$ are given in \AA .}

\label{table:parameters}
\vspace{-3mm}
\begin{center}
  \sisetup{table-format=2.4}

 \begin{tabular}{P{1cm}P{1cm}P{1.8cm}P{1.8cm}P{1.5cm}|P{1cm}P{1cm}P{1.8cm}P{1.8cm}P{1.5cm}}
\hline\hline
 \multirow{1}{*}{} &
 \multicolumn{1}{c}{} &
  \multicolumn{1}{c}{$\mu$ ($m_0$)}  &
  \multicolumn{1}{c}{$\chi_{2D}$ (\AA)} &
  \multicolumn{1}{c|}{$h$ (\AA)} &
\multirow{1}{*}{} &
 \multicolumn{1}{c}{} &
  \multicolumn{1}{c}{$\mu$ ($m_0$)}  &
  \multicolumn{1}{c}{$\chi_{2D}$ (\AA)} &
  \multicolumn{1}{c}{$h$ (\AA)}
   \\ \cline{1-10}
  \hline\
 \multirow{3}{*}{WSe$_2$} & $A_{\text{high}}$ & 0.27 \cite{Ramasubramaniam2012} & 7.18 \cite{Berkelbach2013}  &\multirow{3}{*}{6.575 \cite{Kylanpaa2015}}& \multirow{3}{*}{MoSe$_2$} & $A_{\text{high}}$ & 0.31 \cite{Ramasubramaniam2012} & 8.23 \cite{Berkelbach2013} &  \multirow{3}{*}{6.527 \cite{Kylanpaa2015}}\\

 & $A_{\text{low}}$ & 0.15 \cite{Kormanyos2015} & 7.571 \cite{Kylanpaa2015}  & & & $A_{\text{low}}$ & 0.27 \cite{Berkelbach2013} & 8.461 \cite{Kylanpaa2015} \\
   & $B$ & 0.16 \cite{Kylanpaa2015} & 7.18 \cite{Berkelbach2013} &  & & $B$ & 0.29 \cite{Kylanpaa2015} & 8.23 \cite{Berkelbach2013}  \\ \hline


   \multirow{3}{*}{WS$_2$} & $A_{\text{high}}$ & 0.23 \cite{Ramasubramaniam2012} & 6.03 \cite{Berkelbach2013} &\multirow{3}{*}{6.219 \cite{Kylanpaa2015}} &\multirow{3}{*}{MoS$_2$} & $A_{\text{high}}$ & 0.28 \cite{Ramasubramaniam2012} & 6.60  \cite{Berkelbach2013}&  \multirow{3}{*}{6.18 \cite{Kylanpaa2015}}\\

 & $A_{\text{low}}$ & 0.15 \cite{Kormanyos2015} & 6.393 \cite{Kylanpaa2015} &  & & $A_{\text{low}}$ & 0.16 \cite{Chernikov2014} & 7.112 \cite{Kylanpaa2015} &\\

 & $B$ & 0.15 \cite{Kylanpaa2015}  & 6.03 \cite{Berkelbach2013} & & & $B$ & 0.24 \cite{Kylanpaa2015} & 6.60 \cite{Berkelbach2013} & \\
\hline \hline
\end{tabular}
\end{center}
\vspace{-8mm}
\end{table}

\section{Binding energies of Rydberg excitons in MX$_2$-hBN-MX$_2$ heterostructure}\label{app:rydberg}
\vspace{-25mm}

\begin{table}[H]
\caption{Binding energies of Rydberg excitons in state 1$s$ in van der Waals heterostructures MX$_2$-hBN-MX$_2$. Binding energies are calculated for $A_{\text{high}}$ excitons using V$_{RK}$ and V$_C$. Energy is measured in meV.}

\label{table:binding_energies}
\begin{center}
  \sisetup{table-format=2.4}

 \begin{tabular}{P{1cm}P{1.5cm}P{1.5cm}P{1.5cm}P{1.5cm}P{1.5cm}P{1.5cm}P{1.5cm}P{1.5cm}}
\hline \hline
\multirow{2}{*}{N}&
\multicolumn{2}{c}{WSe$_{2}$-hBN-WSe$_{2}$}&
\multicolumn{2}{c}{WS$_{2}$-hBN-WS$_{2}$} &
\multicolumn{2}{c}{MoSe$_{2}$-hBN-MoSe$_{2}$} &
\multicolumn{2}{c}{MoS$_{2}$-hBN-MoS$_{2}$}
   \\ \cline{2-9}
  &  $V_{RK}$ & $V_C$ & $V_{RK}$ & $V_C$ & $V_{RK}$ & $V_C$& $V_{RK}$ & $V_C$ \\ \cline{1-9}
1 & 107.68 & 243.19 & 110.19 & 223.65 & 105.49 & 260.56 & 113.88 & 247.72\\
2 & 95.61  & 166.95 & 97.16  & 156.13 & 94.12  & 176.36 & 100.30 & 169.42\\
3 & 85.40  & 129.77 & 86.30  & 122.45 & 84.42  & 136.06 & 89.01  & 131.43\\
4 & 77.00  & 107.12 & 77.45  & 101.68 & 76.37  & 111.77 & 79.87  & 108.35\\
5 & 70.04  & 91.68  & 70.20  & 87.39  & 69.68  & 95.32  & 72.37  & 92.65\\
6 & 64.22  & 80.40  & 64.18  & 76.89  & 64.04  & 83.36  & 66.15  & 81.18
\\   \hline \hline

\end{tabular}
\end{center}
\end{table}

\section{Binding energies of magnetoexcitons in MX$_2$-hBN-MX$_2$ heterostructures} \label{app:het}
\begin{figure}[H]
\begin{tabular}{cc}

\textit{(a)} 1$s$ &  \textit{(b)} 2$s$ \\
\
  \includegraphics[width=80mm]{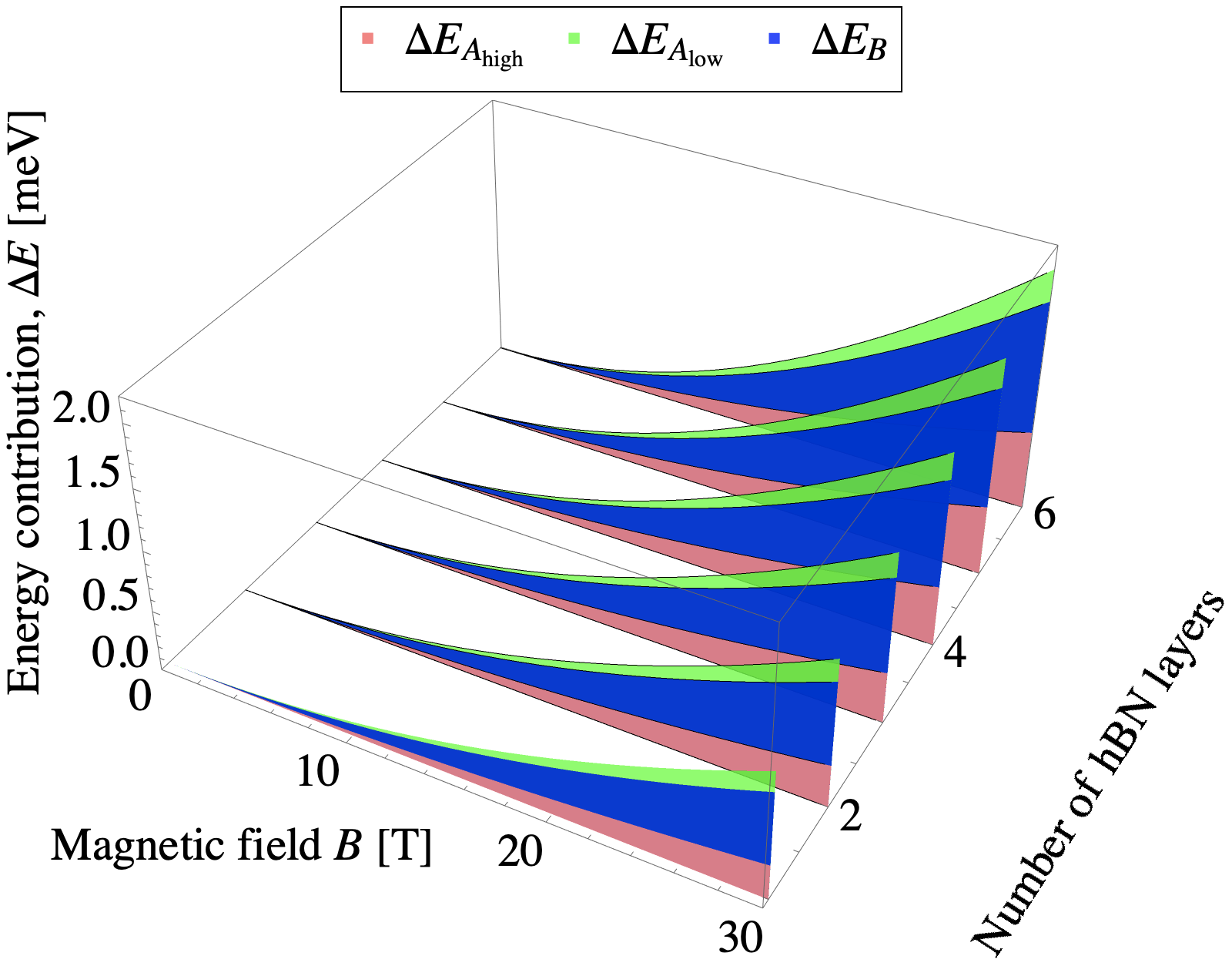} &   \includegraphics[width=80mm]{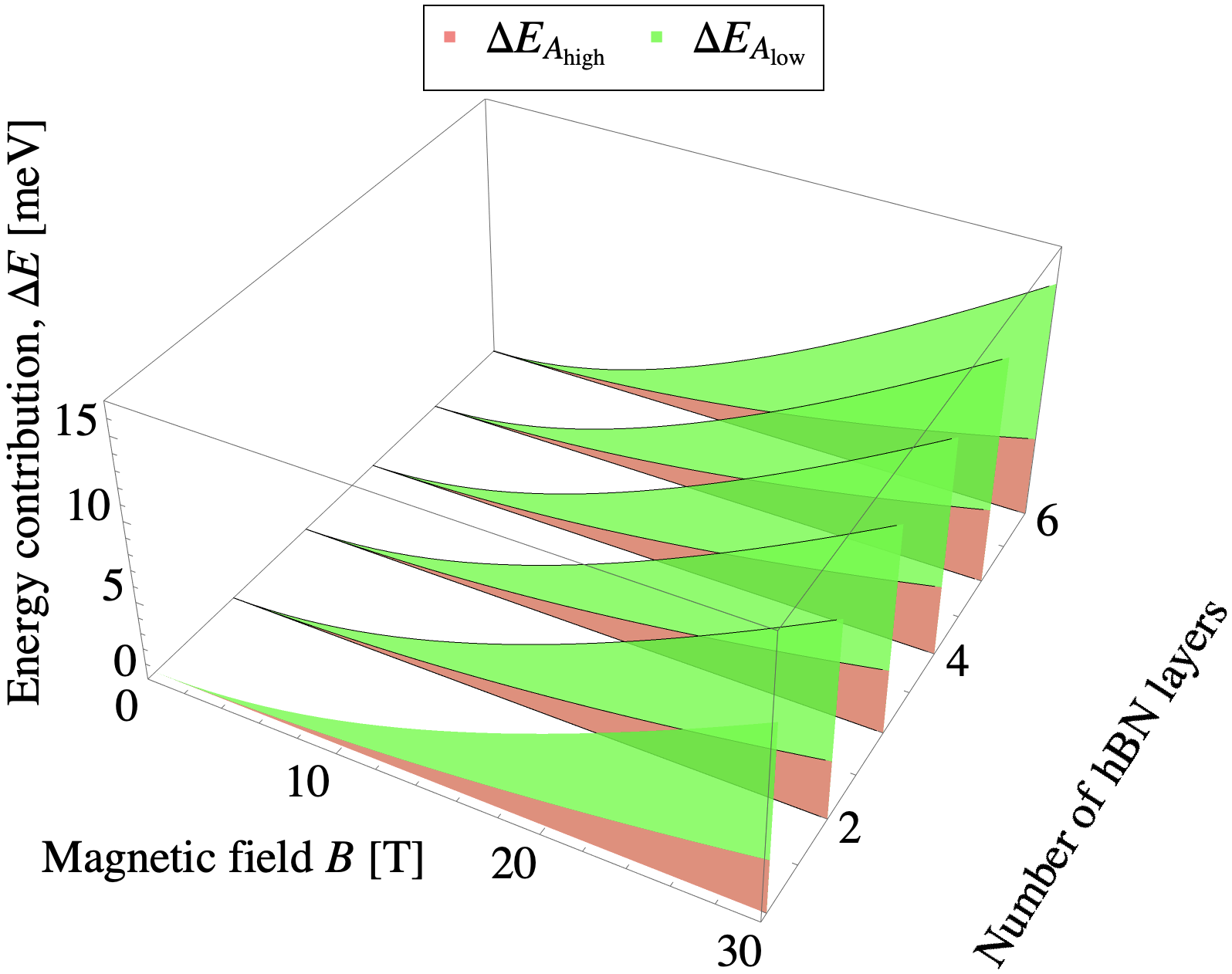} \\[6pt]

\textit{(c)} 3$s$ &  \textit{(d)} 4$s$ \\
\
  \includegraphics[width=80mm]{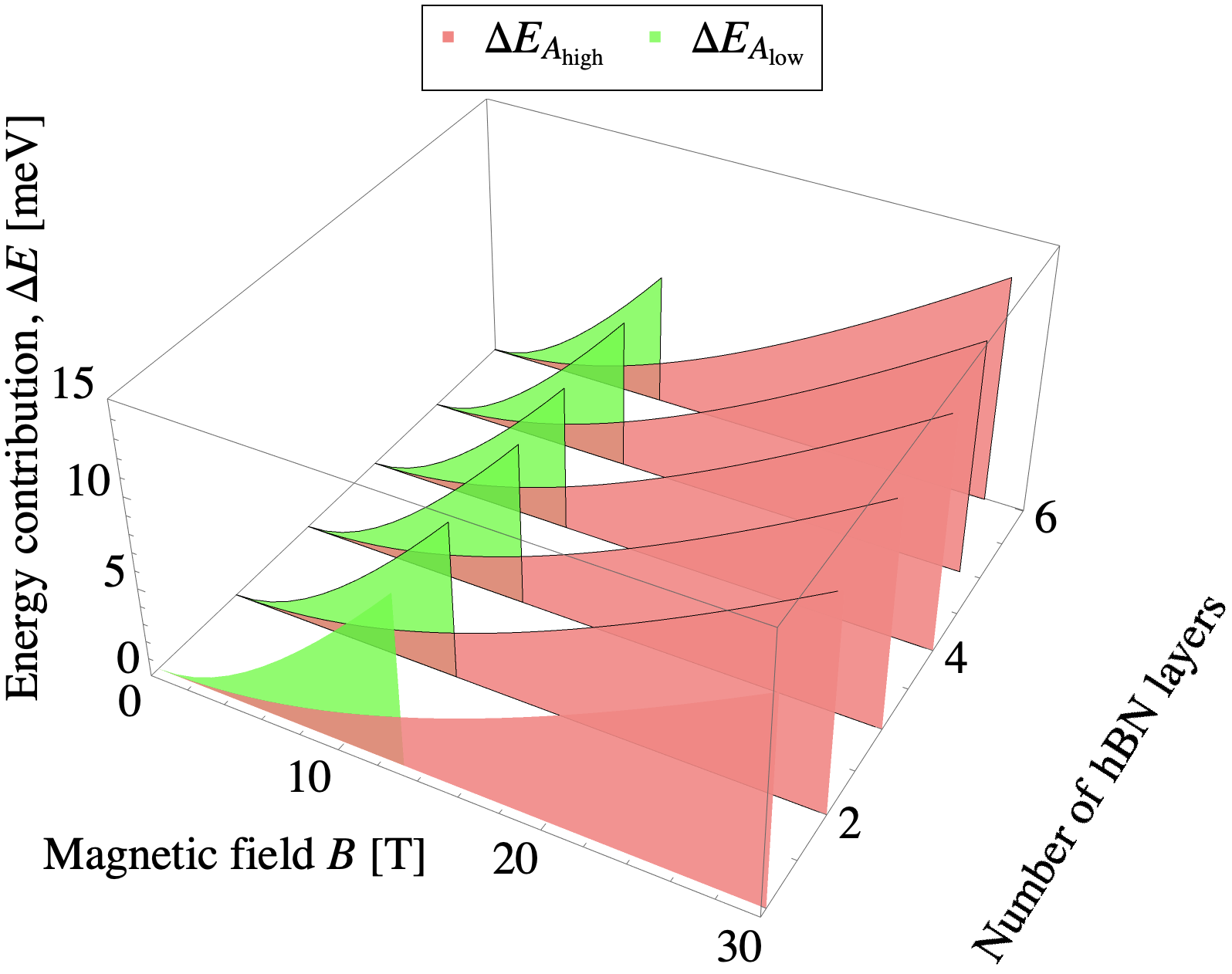} &   \includegraphics[width=80mm]{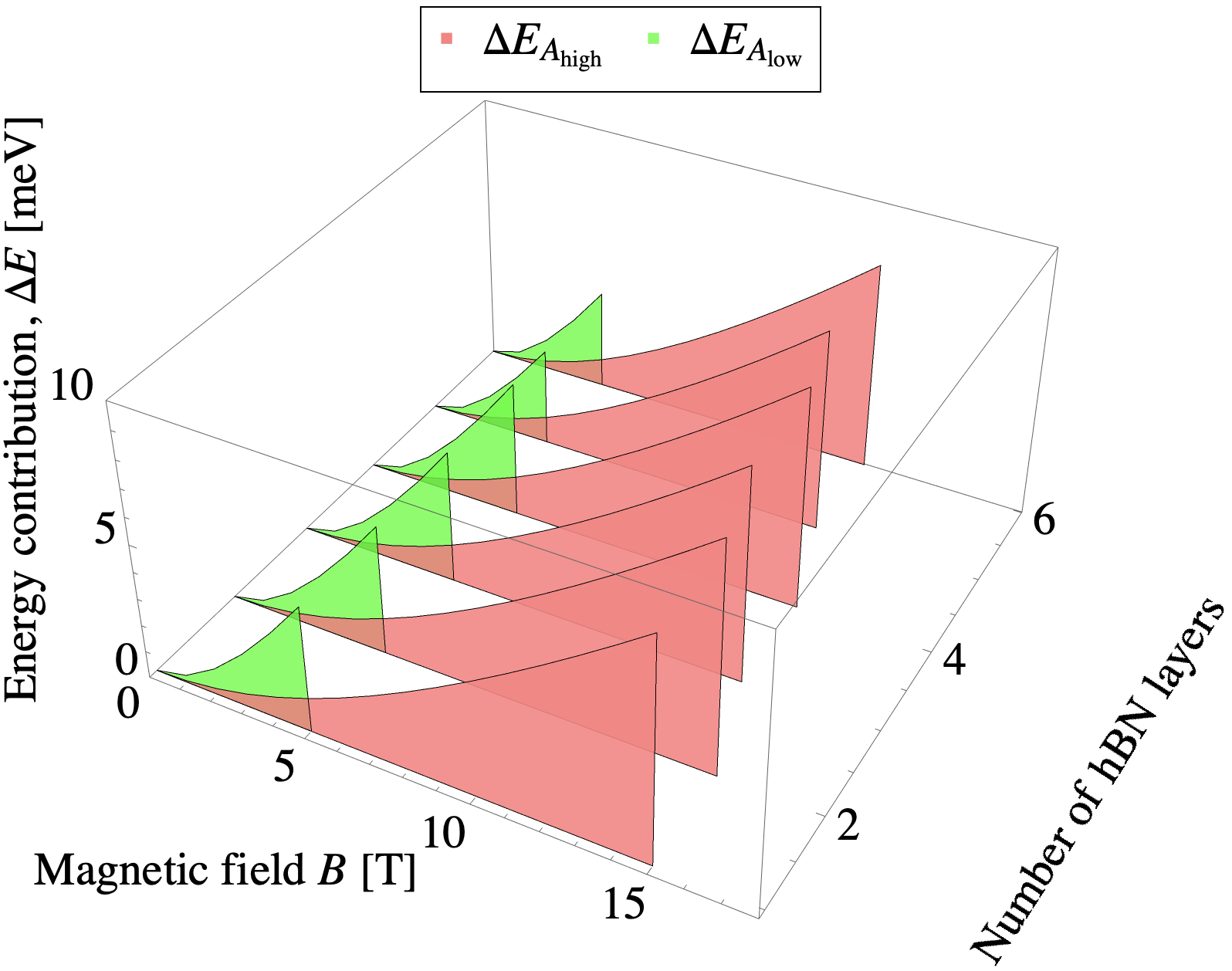} \\[6pt]

\end{tabular}
\caption{The energy contribution from the magnetic field to the binding energies of indirect magnetoexcitons in WSe$_{2}$-hBN-WSe$_{2}$ heterostructure for Rydberg states 1$s$ $(a)$, 2$s$ $(b)$, 3$s$ $(c)$, and 4$s$ $(d)$, respectively. Calculations are performed using the Rytova-Keldysh potential. The surface edge tips for the 3$s$ and 4$s$ state
correspond to the magnetic field where the dissociation of the magnetoexciton occurs.}  \label{wse_rk}
\end{figure}

In Fig. \ref{wse_rk} the energy contributions from the magnetic field to the binding energies of indirect magnetoexcitons in tangerine-based WSe$_2$-hBN-WSe$_2$ heterostructure are presented. The comparison with the results with the indirect magnetoexcitons in MoS$_2$-hBN-MoS$_2$ heterostructure presented in Fig. \ref{mos_rk} shows that the differences occur only in the energy contribution magnitudes, which varies within up to 40\% for MoSe$_{2}$-hBN-MoSe$_{2}$ for the maximum value, and indirect magnetoexcitons do not dissociate in the 3$s$ state and dissociate at larger values of the magnetic field in 4$s$ state.

\begin{figure}[H]
\begin{tabular}{cc}
\textit{(a)} 1$s$ & \textit{(b)} 2$s$ \\
\includegraphics[width=80mm]{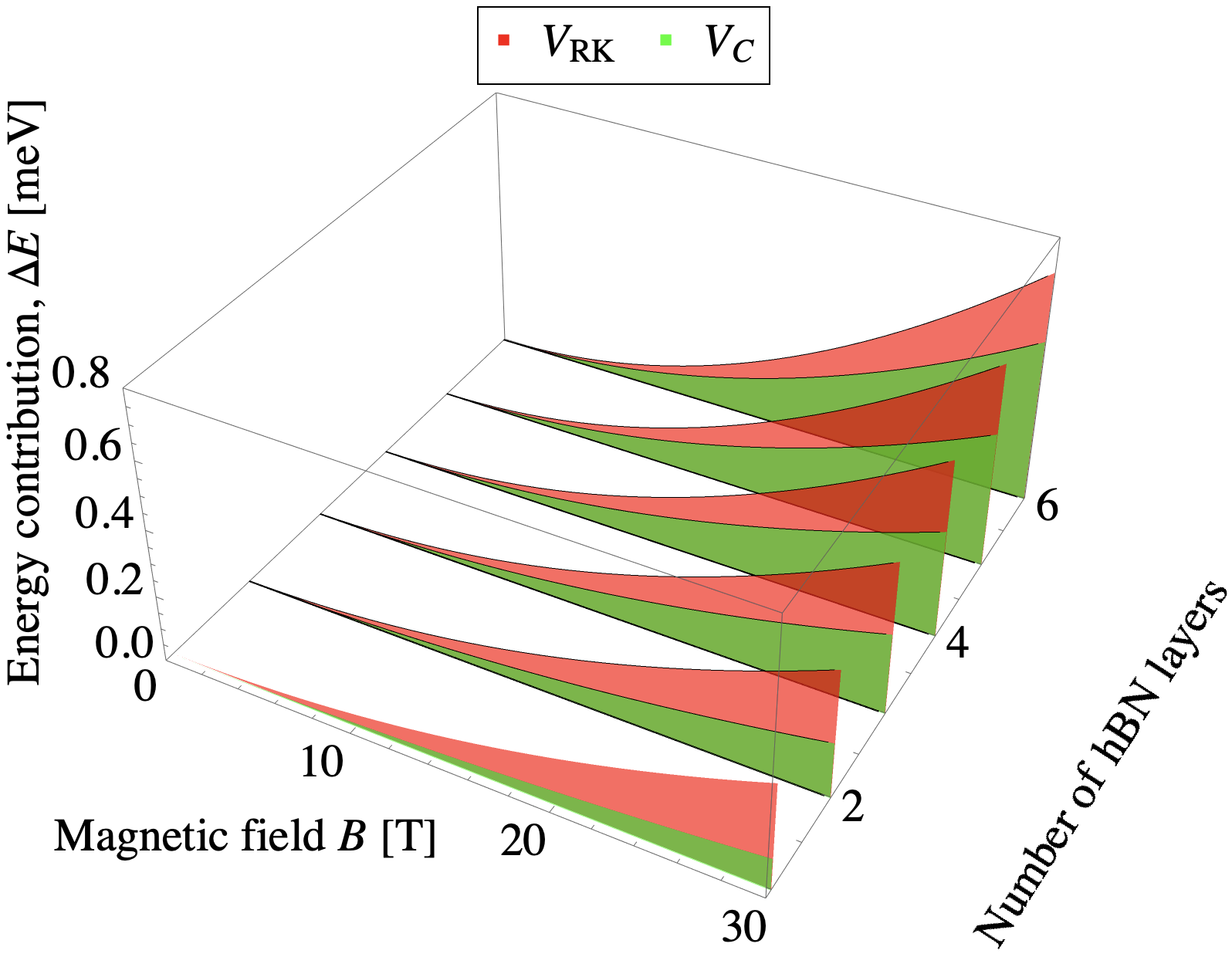} & \includegraphics[width=80mm]{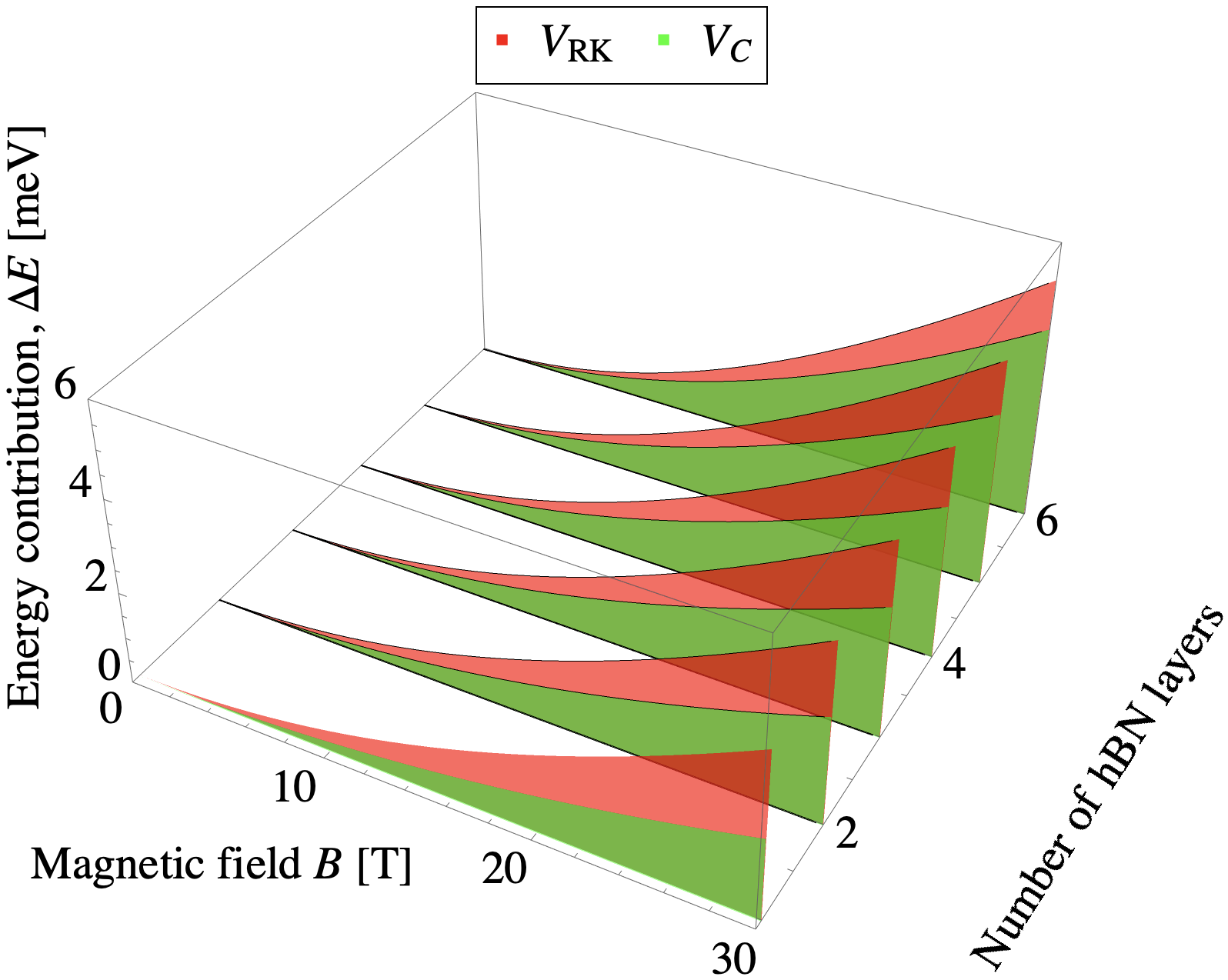}
\\[6pt]
\textit{(c)} 3$s$ & \textit{(d)} 4$s$ \\
\includegraphics[width=80mm]{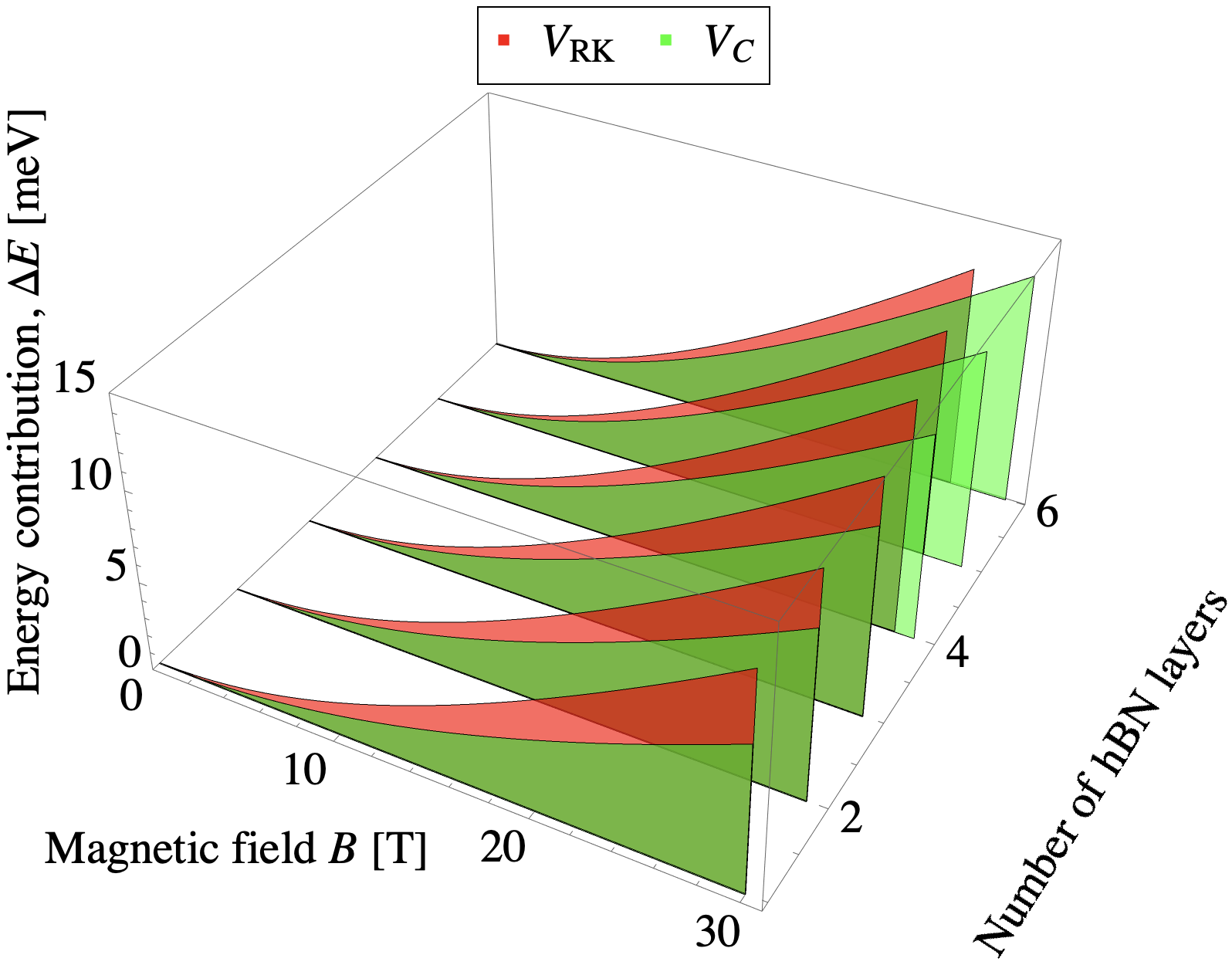} & \includegraphics[width=80mm]{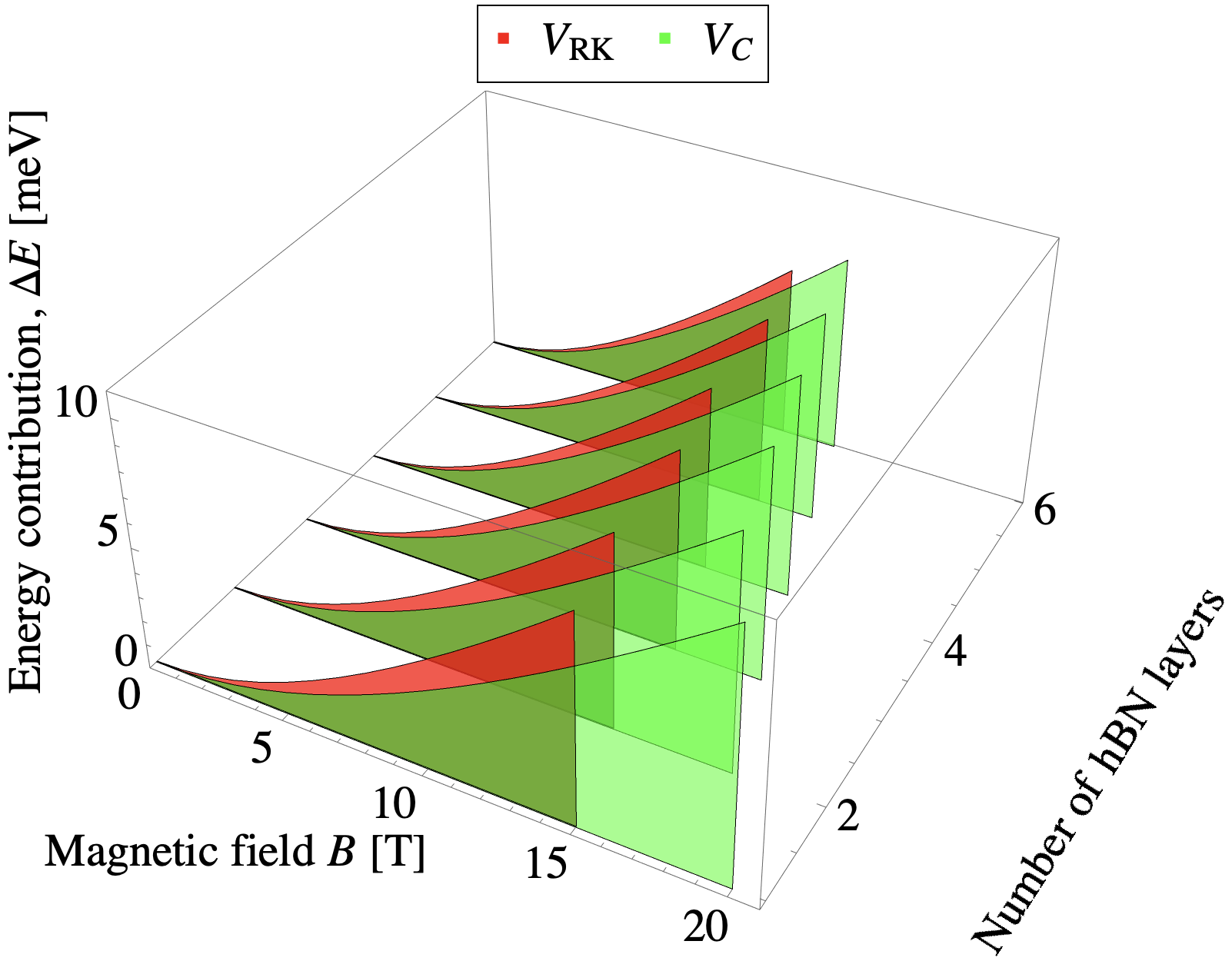}
\\[6pt]
\end{tabular}
\caption{Comparison between energy contribution when Schr\"{o}dinger equation is solved with the RK and Coulomb potentials for indirect $A_{\text{low}}$ magnetoexcitons in MoSe$_2$-hBN-MoSe$_2$ heterostructure for Rydberg states 1$s$ $(a)$, 2$s$ $(b)$, 3$s$ $(c)$, and 4$s$ $(d)$,  respectively. The surface edge tips for the 4$s$ state
correspond to the magnetic field where the dissociation of the magnetoexciton occurs.}
\label{mose_low}
\end{figure}
\section{Diamagnetic coefficients}\label{app:coef}

\begin{table}[H]
\caption{The diamagnetic coefficients of indirect magnetoexcitons in the tungsten-based van der Waals heterostructures. DMCs are calculated using V$_{RK}$ and V$_C$ potentials. The DMCs are obtained when $R^2 = 0.9998$ for the linear regression model. $\sigma$ is given in $\mu$eV/T$^{2}$.}

\label{table:coefficients_S}
\begin{center}
  \sisetup{table-format=2.4}

 \begin{tabular}{P{1cm}P{1cm}P{1cm}P{1cm}P{1cm}P{1cm}P{1cm}P{1cm}|P{1cm}P{1cm}P{1cm}P{1cm}P{1cm}P{1cm}}
\hline \hline
\multicolumn{2}{c}{}&
\multicolumn{6}{c|}{ WSe$_{2}$-hBN-WSe$_{2}$}&
\multicolumn{6}{c}{WS$_{2}$-hBN-WS$_{2}$}
 \\ \cline{3-14}
 \multirow{2}{*}{State} &
 \multirow{2}{*}{$N$} &
 \multicolumn{2}{c}{$\sigma_{A_{\text{high}}}$} &
  \multicolumn{2}{c}{$\sigma_{A_{\text{low}}}$}  &
  \multicolumn{2}{c|}{$\sigma_B$} &
 \multicolumn{2}{c}{$\sigma_{A_{\text{high}}}$} &
  \multicolumn{2}{c}{$\sigma_{A_{\text{low}}}$}  &
  \multicolumn{2}{c}{$\sigma_B$}
   \\ \cline{3-14}
  &&  $V_{RK}$ & $V_C$ & $V_{RK}$ & $V_C$ & $V_{RK}$ & $V_C$ &$V_{RK}$ & $V_C$ & $V_{RK}$ & $V_C$ & $V_{RK}$ & $V_C$
  \\ \cline{1-14}
    \multirow{6}{*}{1$s$}& 1 & 0.32 & 0.11 & 1.16 & 0.43 & 0.98 & 0.37 & 0.40 & 0.16 & 1.05 & 0.43 & 1.02 & 0.43\\
    & 2 & 0.39 & 0.18 & 1.35 & 0.67 & 1.15 & 0.58 &   0.50 & 0.26 & 1.25 & 0.67 & 1.21 & 0.67  \\
    & 3 & 0.47 & 0.27 & 1.57 & 0.93 & 1.34 & 0.81 &   0.60 & 0.38 & 1.46 & 0.93 & 1.43 & 0.93 \\
    & 4 & 0.56 & 0.36 & 1.79 & 1.18 & 1.55 & 1.04 &   0.71 & 0.50 & 1.68 & 1.18 & 1.65 & 1.18 \\
    & 5 & 0.65 & 0.46 & 2.02 & 1.44 & 1.75 & 1.27 &   0.83 & 0.62 & 1.92 & 1.44 & 1.88 & 1.44 \\
    & 6 & 0.74 & 0.55 & 2.25 & 1.71 & 1.97 & 1.51 &   0.95 & 0.75 & 2.15 & 1.71 & 2.12 & 1.71\\ \cline{1-14}

    \multirow{2}{*}{2$s$}&  1 &  & 2.00 &  &  &  &   &  & 3.00	\\
    & 2 &  &  2.65 &  & \\
      \hline \hline

\end{tabular}
\end{center}
\end{table}


\bibliography{/Users/Nastia/Desktop/dissertation/bibliography/bibl_v4}
\end{document}